\documentclass[twocolumn]{aastex631}
\usepackage{comment}
\usepackage{amsmath}
\usepackage{mathrsfs}
\usepackage{graphicx}
\usepackage{booktabs} 
\usepackage{tablefootnote}
\usepackage{longtable}
\usepackage{mhchem}
\usepackage{soul}
\usepackage{xcolor}

\usepackage{tikz}
\usepackage{etoolbox}
\robustify{\tikz}

\begin{document}


\title{A Comprehensive Sulfur Chemistry Network Including Excited \ce{S(^1D)} and \ce{SO(^1\Delta)} for the \texttt{XODIAC} Photochemical Model: Accounting for Missing Sulfur Processes in Venus and Exo-Venus Analogs}

\correspondingauthor{Liton Majumdar}
\email{liton@niser.ac.in, dr.liton.majumdar@gmail.com}

\author[ 0009-0002-4995-9346]{Priyankush Ghosh}
\affiliation{Exoplanets and Planetary Formation Group, School of Earth and Planetary Sciences, National Institute of Science Education and Research, Jatni 752050, Odisha, India}
\affiliation{Homi Bhabha National Institute, Training School Complex, Anushaktinagar, Mumbai 400094, India}

\author[0000-0001-8020-0915]{Namrata Rani}
\affiliation{Exoplanets and Planetary Formation Group, School of Earth and Planetary Sciences, National Institute of Science Education and Research, Jatni 752050, Odisha, India}
\affiliation{Homi Bhabha National Institute, Training School Complex, Anushaktinagar, Mumbai 400094, India}
\affiliation{Departamento de Físico-Química, Facultad de Ciencias Químicas, Universidad de Concepción, Concepción, Chile}

\author[0000-0002-1551-2610]{Jeehyun Yang}
\affiliation{Department of Astronomy and Astrophysics, The University of Chicago, Chicago, IL 60637, USA}

\author[0000-0001-6124-5974]{Karen Willacy}
\affiliation{Jet Propulsion Laboratory, California Institute of Technology, 4800 Oak Grove Drive, Pasadena, CA 91109, USA}

\author[0000-0002-7180-081X]{P. B. Rimmer}
\affiliation{Cavendish Laboratory, University of Cambridge, JJ Thomson Ave, Cambridge CB3 0HE, UK}

\author[0000-0001-7031-8039]{Liton Majumdar}
\affiliation{Exoplanets and Planetary Formation Group, School of Earth and Planetary Sciences, National Institute of Science Education and Research, Jatni 752050, Odisha, India}
\affiliation{Homi Bhabha National Institute, Training School Complex, Anushaktinagar, Mumbai 400094, India}

\begin{abstract}

Sulfur chemistry plays a central role in controlling the atmospheric structure, cloud formation, and composition of Venus and Venus-like exoplanets. However, key reactions involving ground- and excited-state sulfur species remain poorly constrained, and existing photochemical models often rely on incomplete or uncertain kinetic data under high-temperature, \ce{CO2}-rich conditions. In this work, we compute kinetic parameters for reactions of ground-state \ce{S(^3P)} and excited-state \ce{S(^1D)} with \ce{CO2} under Venus-like conditions, forming \ce{SO(^3\Sigma)}, \ce{SO(^1\Delta)}, and \ce{CO}. We characterize the underlying potential energy surfaces, identify intermediate complexes, and derive temperature-dependent rate coefficients using a master-equation framework based on the chemically significant eigenvalue method. We also provide NASA 7-term polynomial coefficients for \ce{S} and \ce{SO} in both ground and excited states to enable consistent incorporation into photochemical models. Incorporating these reactions into the one-dimensional photochemical model \texttt{XODIAC} for Venus produces only minor effects above 60~km due to competing pathways. While the model reproduces most observed sulfur species, discrepancies remain for \ce{S3} and \ce{S4}. Introducing a 1~ppm near-surface atomic sulfur source, representing unresolved deep-atmosphere or surface processes, enhances \ce{S3} and \ce{S4} abundances by 1--2 orders of magnitude and improves agreement with observations. For exo-Venus analogs, the updated chemistry produces modest changes under isothermal conditions. In contrast, in strongly irradiated atmospheres with a high-altitude isotherm and a near-surface sulfur source, it leads to pronounced changes in most sulfur-bearing species, along with significant enhancements in \ce{S(^1D)} and \ce{SO(^1\Delta)}. These results highlight the importance of identifying missing sulfur processes that are not currently included in photochemical models, including pathways leading to excited-state sulfur species and sources of deep-atmosphere sulfur. The inclusion of such processes may improve agreement with observations, as seen for Venus, and has important implications for the chemistry of exo-Venus analogs.

\end{abstract}


\section{Introduction} \label{sec:intro}

Sulfur is the tenth most abundant element in the universe and the fifth most abundant in our solar system \citep{lodders2021relative}. Besides H, C, N, and O, it is one of the most chemically reactive and volatile elements. Its chemical affinity for carbon and oxygen, along with its variable oxidation states, makes it a versatile element of interest in astrophysical environments and planetary atmospheres \citep{titov2018clouds, moses1995post, irwin1999cloud, moses2002photochemistry}. Some studies have also proposed sulfur as a proxy for metallicity in these regions \citep{kama2019abundant, Turrini_2021}. In terrestrial planets of our solar system, such as Earth and Venus, sulfur occurs in diverse chemical forms, ranging from elemental sulfur to sulfur oxides, sulfur polyoxides, sulfuric acid, and its clusters \citep{malin1997sulphur, zhang2012sulfur, titov2018clouds}. These sulfur-bearing species are expected to influence cloud and haze formation in planetary atmospheres. Furthermore, sulfur chemistry has been proposed as a potential trigger for Earth's first Snowball Earth event, while sulfur isotopic ratios serve as key tracers of the Great Oxidation Event \citep{farquhar2000atmospheric, pavlov2002mass, ono2003new, Macdonald_2017, hodgskiss_2019}. Sulfur is also present on other planetary bodies in our solar system, such as Jupiter's moon Io \citep{spencer2000discovery, moses2002photochemistry} and comet 67P/C-G \citep{Calmonte2016}. In addition, the gas giants of our solar system host a wide diversity of sulfur-bearing species, including sulfur oxides, carbon disulfide, and hydrogen sulfide, which play significant roles in shaping their atmospheric compositions \citep{moses1995post, he2020sulfur, zahnle_et_al...1995}. The recent detection of photochemically produced \ce{SO2} in the atmosphere of the gas-giant exoplanet WASP-39b \citep{jwst2023identification, feinstein2023early, tsai2023photochemically} and H$_2$S in HD 189733b \citep{fu2024h2s} using JWST further highlights the importance of sulfur chemistry, not only in the gas giants of our solar system but also in exoplanetary systems.

The role of sulfur in the atmosphere of Venus is critical for understanding its atmospheric evolution and its potential as a natural laboratory for exoplanetary science \citep{Kane19}. The Venusian atmosphere contains thick cloud layers primarily composed of sulfuric acid condensates, formed through the photochemical processing of \ce{SO2} in the presence of a \ce{CO2}-dominated atmosphere \citep{Dai_2024,Rimmer_2021}. These clouds, which regulate the planetary climate, may have transformed early temperate surface conditions into the current harsh and extreme environment by following evolutionary pathways distinct from those of Earth \citep{Dai2025}. Volcanic degassing, as suggested by recent Magellan observations \citep{Herrick_2023}, may play a key role in shaping deep atmospheric chemistry by releasing sulfur-bearing species such as \ce{H2S} and \ce{SO2}. The extreme conditions created by the runaway greenhouse effect offer a natural laboratory for studying high-temperature sulfur chemistry, with implications not only for Venus but also for Venus-like exoplanets \citep{Kane19}. The evaporation–condensation cycle of sulfur is closely coupled with vertical mixing and can affect other reactive species through radical production \citep{zahnle2016photolytic, hobbs2021sulfur, Rimmer_2021}. Moreover, sulfur's ability to alter the concentrations of abundant CO$_2$ and other major atmospheric species may introduce variations in the interpretation of abundances in models of atmospheres that previously excluded sulfur chemistry \citep{veillet2024extensively}.

Due to the high reactivity of sulfur, conducting experiments under astrophysical conditions is challenging. Only a limited number of laboratory studies have incorporated sulfur \citep{he2018gas, he2020sulfur, horst2018haze, dewitt2010formation, farquhar2001observation, reed2020impact, reed2022trace, reed2024abiotic}. Consequently, investigations of sulfur chemistry in planetary atmospheres rely primarily on computational approaches. Incorporating sulfur into atmospheric chemistry models has substantially advanced our understanding of planetary environments, ranging from simple chemical equilibrium models with limited molecular species \citep[e.g.,][]{Blecic_2016, molliere2017, Woitke2017, Stock2018, deka2025nexotrans} to complex photochemical kinetic models that include hundreds of molecular species and detailed gas-phase chemical networks with photochemistry and two- and three-body neutral-neutral processes \citep{allen_1981, moses2011disequilibrium, Venot2012, Rimmer_2016, Rimmer_2021, Tsai_2017, Ranjan20, tsai2021comparative, hobbs2021sulfur, KRASNOPOLSKY2007, KRASNOPOLSKY2012, yang2024chemical, bello2025evidence, hu2025water, Ghosh_2026}. Several of these models explicitly incorporate sulfur chemistry \citep{KRASNOPOLSKY2007, KRASNOPOLSKY2012, Ranjan20, tsai2021comparative, Rimmer_2021, yang2024chemical, hu2025water, Wogan25, Ghosh_2026}. Typically, three major classes of sulfur reactions are included in these networks: (i) oxidation pathways involving OH radicals that produce oxides or sulfuric acid, (ii) reactions with hydrocarbons that form CS$_n$ or OCS, and (iii) polymerization of S into S$_n$ \citep{he2020sulfur}. These reaction networks are generally constructed using Arrhenius-type rate coefficients derived from laboratory studies or \textit{ab initio} quantum chemical calculations.

However, current sulfur reaction networks require substantial improvement to accurately describe atmospheric sulfur chemistry. This need arises from the absence of experimentally measured or quantum chemically computed kinetic data across wide temperature ranges, as well as the omission of several key reactions and processes \citep{hobbs2021sulfur}. For example, carbon dioxide is a primary species in both thermochemical equilibrium and photochemical models \citep{woitke2021coexistence, liggins2023atmospheric, petralia2020systematic} and serves as the dominant carbon reservoir in the Venusian atmosphere. It is also widely used as an indicator of both primary and secondary planetary atmospheres \citep{swain2021detection}. Modeling studies of gas giants suggest that sulfur can reduce \ce{CO2} abundances by up to five orders of magnitude \citep{moses2016composition, zahnle2016photolytic}, implying a strong coupling between sulfur chemistry and the observability of \ce{CO2}. Despite the importance of this reaction, neither experimentally measured nor \textit{ab initio} derived rate coefficients are available. In the absence of such data, very low rate coefficients of order 10$^{-20}$ cm$^3$\,molecule$^{-1}$\,s$^{-1}$ have been adopted in a few chemical networks, including \texttt{EPACRIS} \citep{Yang2024epacris} and \texttt{KINETICS} \citep{willacy2022vertical}. These assumptions may lead to inaccurate predictions of sulfur-bearing species in photochemical models of planetary atmospheres.

In addition, reactions involving excited-state sulfur S($^1$D), alongside ground-state sulfur \ce{S(^3P)}, are often neglected. Long-lived excited species can be significantly more reactive than their ground-state forms, and prior work on photoexcited CO \citep{Yang_2023} shows how photoexcitation can introduce fast, otherwise inaccessible pathways that reshape the overall chemistry under photon-rich conditions. For this reason, a recent study by \citet{Veillet2025arxiv} incorporated \ce{S(^1D)} to explore carbon–sulfur coupling in exoplanetary atmospheres, but excited-state sulfur chemistry remains largely absent from most photochemical models. Missing species such as S($^1$D) and \ce{SO(^1\Delta)}, their associated thermochemistry, accurate sulfur--\ce{CO2} reaction kinetics, and newly identified photochemical pathways capable of generating these states are crucial components that need to be incorporated into future chemical networks. Including these processes is essential for quantifying their roles in Venus, where at least half a dozen sulfur-bearing species have been detected, and for assessing their implications for potentially rocky, Venus-like exoplanets.

In this study, we address these gaps by introducing the reaction between atomic sulfur (in both its ground and excited states) and carbon dioxide using high-level quantum chemical simulations. We derive kinetic parameters for these fundamental reactions and incorporate them into the \texttt{XODIAC-2025} chemical network \citep{Ghosh_2026}. In addition, we expand the network by proposing new reactions involving \ce{S(^1D)}, $\mathrm{SO}(^1\Delta)$, and ground-state SO, along with additional photochemical formation pathways for these species. We extend the thermochemical database by computing NASA 7-term polynomial coefficients for \ce{S(^1D)} and $\mathrm{SO}(^1\Delta)$. Using this updated \texttt{XODIAC-2025} network and thermochemical database, we employ the state-of-the-art one-dimensional photochemical model \texttt{XODIAC} \citep{Ghosh_2026} to investigate the impact of the proposed chemistry on Venus and three exo-Venus analogs. For Venus, we compare our results with observationally constrained sulfur abundances and examine how well the extended network reproduces the measured abundances of the multiple sulfur-bearing species detected to date.

This paper is organized as follows. Section~\ref{sec:method} describes the computational methodology used to obtain the electronic structure, kinetic, and thermochemical data for the proposed sulfur chemistry, as well as the photochemical kinetics model employed in this work (Section~\ref{sec:photomodeling_section}). Section~\ref{sec:results} presents the quantum chemical results, including reaction profile analyses, temperature-dependent rate coefficients, new thermochemical parameters, and their astrophysical implications for Venus and the three exo-Venus analog atmospheres. Finally, Section~\ref{sec:conclusions} summarizes the main findings of this study.


\section{Computational Methods for Extending the \texttt{XODIAC} Chemical Network and Thermochemical Database}\label{sec:method}

\subsection{Electronic structure calculations} \label{sec:electronic}

The reactant system under consideration is \ce{[S + CO2]}, where \ce{CO2} is in its ground electronic state \(({}^{1}\Sigma)\), and atomic sulfur is examined in both the ground state \ce{S(^3P)} and the first excited state \ce{S(^1D)}. To investigate the reaction mechanism, preliminary transition-state (TS) searches were carried out at the B3LYP-D3(BJ)/6-311++G(2d,2p) level of theory \citep{becke1993density, lee1988development, krishnan1980self}. The geometries of all stationary points located were further refined at the WB97X-D/def2-TZVP level of theory \citep{chai2008long, weigend2005balanced}. Harmonic frequency analyses were performed to obtain zero-point vibrational energies (ZPEs) and to characterize local minima (no imaginary frequencies) and transition states (one imaginary frequency). For each TS, intrinsic reaction coordinate (IRC) calculations were carried out \citep{fukui1981path} at the WB97X-D/def2-TZVP level. Single-point electronic energies were further refined at the CCSD(T)/aug-cc-pV(T+d)Z level of theory \citep{raghavachari1989fifth} using the WB97X-D/def2-TZVP geometries. Notably, all DFT calculations were carried out using restricted open-shell formalisms and single-point energy refinements were performed using ROCCSD(T), based on restricted open-shell Hartree–Fock reference wavefunctions. Further, to assess the possible role of static correlation, we evaluated the T$_1$ diagnostic for all stationary points \citep{lee1988development}; the resulting values ranging between 0.019-0.029 indicate predominantly single-reference character for both singlet and triplet species \citep{ali2018diamine, klippenstein2022}, reported in Table \ref{tab:T1_diagnostics} of Appendix. All electronic structure calculations were performed with \textsc{Gaussian}~16, Revision~A.03 \citep{G16A03}. Finally, the fine-structure spin-orbit splitting of ground-state atomic sulfur \ce{S(^3P)}, which is not captured in non-relativistic electronic structure calculations, was accounted by applying spin-orbit correction that lowers the \ce{S(^3P)} asymptotic energy by 2.3~kJmol$^{-1}$, based on the experimentally determined fine-structure splitting from NIST Atomic Spectra Database \citep{NISTDatabase}. Moreover, the energy of the \ce{S(^1D)} electronic state was obtained by applying the experimentally determined \ce{S(^3P)}–\ce{S(^1D)} excitation energy (110.5~kJmol$^{-1}$) as \textit{a posteriori} correction to the computed \ce{S(^3P)} reference energy at all the levels of theory as implemented by \cite{di2025hot}. An analogous procedure was adopted for the SO molecule, where the energy of the singlet state \ce{SO(^1\Delta)} was obtained by adding 77.1~kJmol$^{-1}$ to the computed values of \ce{SO(^3\Sigma)}\citep{borin1999lowest}, ensuring a consistent and experimentally anchored description of the relative energies of atomic and molecular singlet excited states.

\subsection{Kinetic calculations} \label{sec:kinetic}

The rate coefficients for all reaction pathways were computed using the Master Equation System Solver (\texttt{MESS}) code \citep{georgievskii2013reformulation} based on the potential energy surfaces calculation described in Section~\ref{sec:electronic}. \texttt{MESS} is based primarily on Master Equation methods \citep{miller2006master}, employing phase space theory (PST) to describe barrierless reactions and transition state theory (TST) for channels involving one or more transition states (TSs) \citep{bao2017variational}.  

Carbon dioxide was treated as the buffer gas, and the Lennard--Jones (LJ) parameters of the reactive system were taken from the literature for the analogous \ce{CO2}/O gas-phase system \citep{li2012vapor}. For the barrierless channel, the long-range interaction potential was modeled as a $-C_n/R^n$ power law, where $R$ is the distance between the two approaching fragments (in Bohr). In this case, $n = 6$, corresponding to the isotropic dispersion interaction \citep{fernandez2006modeling}. To obtain the potential prefactor $C_6$, a relaxed scan of the interaction curve was performed by varying the distance between sulfur and the oxygen atom of \ce{CO2} at the ROWB97X-D/def2-TZVP followed by single-point energy calculations refined at the ROCCSD(T)/aug-cc-pV(T+d)Z level of theory, Figure~\ref{fig:6}, in Appendix. The interaction energies in the 3.5-7.0{~\AA}  are then fitted to the $-C_6/R^6$ form, yielding $C_6 = 93.63$ a.u.

Rate coefficients were calculated at a pressure of 1~bar over the temperature range 150--2000~K using the direct diagonalization approach implemented in \texttt{MESS} \citep{georgievskii2013reformulation}. These values correspond to the overall reaction pathway, rather than stepwise elementary processes. The temperature-dependent rate coefficients were then fitted to the Modified Arrhenius equation \citep{laidler1996glossary}:
\begin{equation}
k(T) = \alpha \left(\frac{T}{300 K}\right)^{\beta} \exp\left(-\frac{\gamma}{T}\right),
\
\label{eq:arrhenius_eqn}
\end{equation}
where $\alpha$ ($\mathrm{cm^3\,molecule^{-1}\,s^{-1}}$), $\beta$, and $\gamma$ (K) are the fitted parameters.

\begin{table*}[ht!]
 \caption{Summary of all Venusian and exo-Venusian models considered in this study, including variations in the T--P profile, stellar flux, chemical network, and atomic sulfur abundance.}
 \label{tab:model_summary}
 \centering
 \small
 \setlength{\tabcolsep}{6pt}
 \begin{tabular}{|c|c|c|c|c|c|}
 \hline
 Model & T-P Profile & Incident Flux at TOA & Chemical Network & S mixing ratio & Model Type \\
 \hline
 M0 & \citet{KRASNOPOLSKY2007} & 1$\times$solar  & \texttt{XODIAC-2025.v1} & 0 & Venusian \\
 \hline
 M1 & \citet{KRASNOPOLSKY2007} & 1$\times$solar  & \texttt{XODIAC-2025.v1} & 1~ppm & Venusian \\
 \hline
 M2 & \citet{KRASNOPOLSKY2007} & 1$\times$solar  & \texttt{XODIAC-2025.v2} & 0 & Venusian \\
 \hline
 M3 & \citet{KRASNOPOLSKY2007} & 1$\times$solar  & \texttt{XODIAC-2025.v2} & 1~ppm & Venusian \\
 \hline
 M4 & \citet{KRASNOPOLSKY2007} + 505.6 K isotherm & 1$\times$solar  & \texttt{XODIAC-2025.v2} & 0 & exo-Venusian \\
 \hline
 M5 & \citet{KRASNOPOLSKY2007} & 1000$\times$solar  & \texttt{XODIAC-2025.v2} & 0 & exo-Venusian \\
 \hline
 M6 & \citet{KRASNOPOLSKY2007} + 505.6 K isotherm & 1000$\times$solar  & \texttt{XODIAC-2025.v2} & 0 & exo-Venusian \\
 \hline
 M7 & \citet{KRASNOPOLSKY2007} + 505.6 K isotherm & 1$\times$solar  & \texttt{XODIAC-2025.v2} & 1 ppm & exo-Venusian \\
 \hline
 M8 & \citet{KRASNOPOLSKY2007} & 1000$\times$solar  & \texttt{XODIAC-2025.v2} & 1 ppm & exo-Venusian \\
 \hline
 M9 & \citet{KRASNOPOLSKY2007} + 505.6 K isotherm & 1000$\times$solar  & \texttt{XODIAC-2025.v2} & 1 ppm & exo-Venusian \\
 \hline

 \end{tabular}
\end{table*}

\subsection{Thermochemistry calculations} \label{sec:thermochemistry} 

A critical requirement for assessing the implications of newly computed reactions in photochemical kinetics models for planetary atmospheres is the inclusion of both forward and reverse reactions to establish equilibrium kinetics. Consequently, most photochemical kinetics models in the planetary and exoplanetary community employ NASA polynomial coefficients \citep{burcat_2005} to obtain reverse rates. For the reactions studied here, however, such coefficients describing the thermochemical parameters are not available for S($^1$D) and $\mathrm{SO}(^1\Delta)$. To compute the NASA 7-term polynomial coefficients for these species, we used \texttt{Arkane} (Automated Reaction Kinetics and Network Exploration) \citep{dana_arkane}, a module of \texttt{RMG} \citep[][Reaction Mechanism Generator]{gao2016reaction, liu2021rmg, johnson2022rmg}, which employs thermochemical data extracted from \textsc{Gaussian}~16 \citep{G16A03}. As S($^1$D) and $\mathrm{SO}(^1\Delta)$, the excited states of S and SO, are the singlet and biradical; so we performed broken-symmetry DFT calculations \citep{neese2004definition} using UWB97X-D/def2-TZVP level of theory to obtain a stable single-determinant representation of the target spin state. These calculations were carried out using an unrestricted Kohn-Sham formalism with a mixed initial guess (guess=mix), followed by wavefunction stability analysis (stable=opt). The resulting broken-symmetry singlet solutions exhibit expectation values of the spin operator $\langle S^2 \rangle \approx 1.0$, consistent with a biradical singlet description. For the ground state S and SO the conventional optimization and frequency calculations were performed at the same level of DFT that is UWB97X-D/def2-TZVP The results were then processed with \texttt{Arkane} \citep{dana_arkane}, which fits NASA 7-term polynomial coefficients to thermodynamic properties, including heat capacity, enthalpy, entropy, and Gibbs free energy, over the temperature range 10--3000 K.

\section{Photochemical Kinetics Modeling with \texttt{XODIAC} and the Extended \texttt{XODIAC} Chemical Network} 
\label{sec:photomodeling_section}

To assess the impact of thermochemical and photochemical sulfur kinetics on Venus and exo-Venus atmospheres, we extended the \texttt{XODIAC-2025.v1} base chemical network from \citet{Ghosh_2026} by adding several new thermochemical and photochemical reactions. We refer to this expanded network as \texttt{XODIAC-2025.v2}, which incorporates both ground- and excited-state S and SO species. We simulated Venus and three exo-Venus analogs using the one-dimensional photochemistry–transport model \texttt{XODIAC} \citep{Ghosh_2026} coupled with the \texttt{XODIAC-2025.v2} network. The newly added reactions, including two-body processes and photochemical pathways, together with their relaxation channels through CO$_2$ and N$_2$, are listed in Table~\ref{tab:merged_reactions}.

For reactions 11 and 12 in Table~\ref{tab:merged_reactions}, we adopted the latest JPL photochemical cross-section data \citep{burkholder2019evaluation19}, interpolating the photodissociation cross sections onto a uniform wavelength grid with 1~\AA\ spacing over 1 to 10,000~\AA. These data were supplemented with \normalsize\rm P\small HIDRATES cross sections \citep{Huebner_Carpenter_1979,Huebner_1992,huebner2015photoionization}. Additionally, four photochemical channels leading to the formation of \ce{S(^1D)} and \ce{SO(^1\Delta)} were incorporated by analogy with the corresponding ground-state photochemistry. The photochemical branching ratios between ground and excited states are estimated using statistical weights based on spin multiplicity, given by \(2S+1\). For cases where both singlet (\(S=0\)) and triplet (\(S=1\)) product channels are accessible, their respective degeneracies are 1 and 3, yielding branching fractions of 25\% for the singlet channel (e.g., \ce{S(^1D)}, \ce{SO(^1\Delta)}) and 75\% for the triplet (ground-state) channel. This assumption reflects a statistical partitioning in the absence of detailed state-resolved cross sections (see Table~\ref{tab:merged_reactions}).

For the kinetic simulations, we modeled Venus and three exo-Venus analogs using the full \texttt{XODIAC-2025.v2} thermochemical and photochemical network. For Venus, we considered four cases:  
(1) Model M0, which adopts the full $\mathrm{T\!-\!P\!-\!K_{zz}}$ profile from \citet{KRASNOPOLSKY2007,KRASNOPOLSKY2012} and employs the default \texttt{XODIAC-2025.v1} network with the lower-boundary mixing ratios from \citet{Rimmer_2021}. The species \ce{CO2}, \ce{N2}, \ce{SO2}, \ce{H2O}, \ce{OCS}, \ce{CO}, \ce{HCl}, \ce{H2}, \ce{H2S}, and \ce{NO} were assigned mixing ratios of 0.96, 0.03, 150~ppm, 30~ppm, 5~ppm, 20~ppm, 500~ppb, 3~ppb, 10~ppb, and 5.5~ppb, respectively. For the stellar spectrum, we used the compiled solar flux at the top of the atmosphere over 1 to 10,000~\AA\ \citep{Matthes2017,coddington2015}, scaled for Sun–Earth and Sun–Venus distance ratios following \citet{Rimmer_2021}.  
(2) Model M1, which is identical to M0 but includes a non-zero lower-boundary atomic sulfur abundance of 1~ppm (see Section~\ref{sec:nonzero_sulfur}).  
(3) Model M2, which is identical to M0 but employs the \texttt{XODIAC-2025.v2} network.  
(4) Model M3, which is identical to M2 but also includes a non-zero lower-boundary atomic sulfur abundance of 1~ppm (see Section~\ref{sec:nonzero_sulfur}).

For the exo-Venus analogs, we considered six scenarios:  
(1) Model M4, which employs the \texttt{XODIAC-2025.v2} network, follows the \citet{KRASNOPOLSKY2007} T–P profile up to approximately 30~km, with an isothermal extension at 505.6~K above this altitude, uses the same stellar spectrum as Models M0 to M3, and zero lower-boundary atomic sulfur abundance.  
(2) Model M5, which is identical to M4 but receives 1000 times the solar insolation.  
(3) Model M6, a hybrid model that uses the \citet{KRASNOPOLSKY2007} T–P profile with a 505.6~K isotherm above approximately 30~km together with 1000 times solar insolation. 
(4) M7, which is identical to M4, but also includes a non-zero lower-boundary atomic sulfur abundance of 1~ppm.
(5) M8, which is identical to M5, but also includes a non-zero lower-boundary atomic sulfur abundance of 1~ppm.
(6) M9, which is identical to M6, but also includes a non-zero lower-boundary atomic sulfur abundance of 1~ppm.
A summary of all models is provided in Table~\ref{tab:model_summary}.

\section{Results} \label{sec:results}

\subsection{New Kinetic and Thermochemical Data and the Extended \texttt{XODIAC-2025.v2} Chemical Network}

\subsubsection{Potential Energy Surfaces of S($^3$P) + CO$_2$ and S($^1$D) + CO$_2$}

\begin{figure}[!h]
\includegraphics[width=\columnwidth]{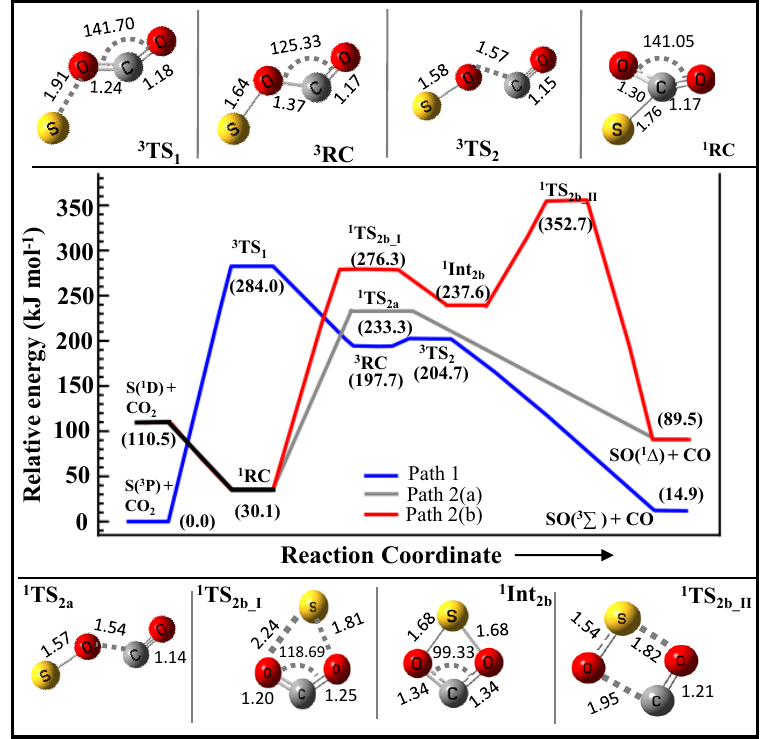}
\caption{Reaction pathways for the S + \ce{CO2} system. The relative potential energy profile (in kJ mol$^{-1}$) was computed at the CCSD(T)/aug-cc-pV(t+d)Z level of theory. Path~1 corresponds to the ground-state triplet surface, while Paths~2(a) and~2(b) correspond to the excited-state singlet surface. Key stationary points are labeled with their relative energies in parentheses. Optimized geometries at the WB97X-D/def2-TZVP level are shown with selected bond lengths (\AA) and bond angles (\textdegree).}

\label{fig:1}
\end{figure}

The potential energy surface (PES) for the studied reaction system is shown in Figure~\ref{fig:1}. We identified three pathways: Path~1 corresponds to \ce{S(^3P)} + \ce{CO2(^{1}\Sigma)}, while Paths~2(a) and 2(b) correspond to \ce{S(^1D)} + \ce{CO2(^{1}\Sigma)}. For Path~1, a large entrance barrier of 284~kJ~mol$^{-1}$ is found relative to the reactants, reflecting the high stability of \ce{CO2} in its ground state. The resulting triplet reaction complex ($^3$RC) is an unstable thionyl formate radical [SOCO], formed via transition state $^3$TS$_1$. The radical $^3$RC then crosses a small energy barrier of 7~kJ~mol$^{-1}$ via transition state $^3$TS$_2$ to yield the products $\mathrm{SO}(^3\Sigma)$ and \ce{CO(^{1}\Sigma)}, where \ce{CO(^{1}\Sigma)} denotes the ground state of CO. Overall, the formation of $\mathrm{SO}(^3\Sigma)$ and \ce{CO(^{1}\Sigma)} proceeds through a two-step process, with $^3$RC formation being the rate-determining step in Path~1. For the remainder of this paper, the ground states of \ce{CO2} and \ce{CO} are denoted simply as \ce{CO2} and \ce{CO}.

Further, owing to the high reactivity of excited-state atomic sulfur and the identical spin multiplicities of the reactants, the formation of the singlet reaction complex in Path~2 occurs without a barrier (Figure~\ref{fig:1}). As singlet sulfur approaches carbon dioxide, its strong nucleophilic character leads to the formation of the reaction complex $^1$RC. Two pathways, Path~2(a) and Path~2(b), emerge from this complex, both yielding $\mathrm{SO}(^1\Delta)$ and \ce{CO} (Figure~\ref{fig:1}). In Path~2(a), $^1$RC dissociates via transition state $^1$TS$_{\rm 2a}$, which has a high energy barrier of 203.2~kJ~mol$^{-1}$. In Path~2(b), $^1$RC rearranges into a cyclic intermediate $^1$Int$_{\rm 2b}$ through transition state $^1$TS$_{\rm 2b\_I}$, requiring an even higher barrier of 246.2~kJ~mol$^{-1}$. This intermediate then dissociates via $^1$TS$_{\rm 2b\_{II}}$, with a barrier of 115.1~kJ~mol$^{-1}$, to yield the products. Both Path~2(a) and Path~2(b) are overall exothermic, although they involve very high activation barriers, partially compensated by the energy released during $^1$RC formation. Additional stabilization may also occur when singlet \ce{SO} relaxes to its more stable triplet ground state.

In addition, to the adiabatic reaction pathways discussed, the reacting system \ce{S(^3P)} + \ce{CO2} may undergo spin crossover to the singlet state \ce{S(^1D)} + \ce{CO2} via a minimum energy crossing point (MECP), as reported for systems involving \ce{O(^3P)} \citep{recio2022intersystem, cavallotti2020theoretical}. However, because the triplet surface exhibits a very large barrier for reaction (Figure \ref{fig:1}), any such crossing would only become relevant at energies comparable to this barrier and is thus expected to be of less importance under low to moderate temperature conditions. We further note that non-adiabatic effects may also influence the fate of the singlet entrance complex \ce{^{1}RC}, which is strongly stabilized with respect to \ce{S(^1D)} + \ce{CO2} (Figure~\ref{fig:1}). In particular, \ce{^{1}RC} could, in principle, undergo intersystem crossing to the triplet spin state via \ce{^{3}RC} and re-dissociate to \ce{S(^3P)} + \ce{CO2} if a nearby singlet-triplet crossing exists. Such a channel could compete with progression of the \ce{^{1}RC} toward \ce{SO(^1\Delta)} + \ce{CO}, especially given the comparatively stronger spin-orbit coupling (SOC) interactions in sulfur-containing systems \citep{maiti2004importance}. Quantifying this competition would require explicit MECP searches and non-adiabatic kinetic treatments, which is beyond the scope of the present work.

\begin{table*}[ht!]
 \caption{Newly computed and proposed two-body and photodissociation reactions added to the \texttt{XODIAC-2025.v1} network, incorporating S($^1$D) and $\mathrm{SO}(^1\Delta)$ chemistry (\texttt{XODIAC-2025.v2} network). The fitted parameters are $\alpha$ ($\mathrm{cm^3\,molecule^{-1}\,s^{-1}}$), $\beta$, and $\gamma$ (K) (Equation \ref{eq:arrhenius_eqn}). Numbers in parentheses denote exponents, with the values on the left representing the corresponding bases.}
 \label{tab:merged_reactions}
 \centering
 \small
 \setlength{\tabcolsep}{6pt}
 \begin{tabular}{|c|c|c|c|c|c|c|}
 \hline
  \# & Reactions & Type & $\alpha$ & $\beta$ & $\gamma$ & Reference\\
 \hline
 1. & $\mathrm{S + CO_2 \rightarrow SO + CO}$ & Two-body  & $4.48(-15)$ & {$1.4$} & $3.33(+4)$ & Path 1 in Figure~\ref{fig:1} \\
 2. & $\mathrm{S + CO_2 \leftarrow SO + CO}$ & Two-body & $2.48(-20)$ & $2.28$ & $3.07(+4)$ & Path 1 in Figure~\ref{fig:1} \\
 3. & $\mathrm{S(^1D) + CO_2 \rightarrow SO(^1\Delta) + CO}$ & Two-body& $1.53(-11)$ & $0.57$ & $1.55(+4)$ & Path 2(a) in Figure~\ref{fig:1} \\
 4. & $\mathrm{S(^1D) + CO_2 \leftarrow SO(^1\Delta) + CO}$ & Two-body& $1.15(-16)$ & $1.43$ & $1.75(+4)$ & Path 2(a) in Figure~\ref{fig:1} \\
 5. & $\mathrm{S(^1D) + CO_2 \rightarrow SO(^1\Delta) + CO}$ & Two-body  & $2.35(-12)$ & $0.6$ & $2.97(+4)$ & Path 2(b) in Figure~\ref{fig:1} \\
 6. & $\mathrm{S(^1D) + CO_2 \leftarrow SO(^1\Delta) + CO}$ & Two-body  & $1.40(-17)$ & $1.49$ & $3.16(+4)$ & Path 2(b) in Figure~\ref{fig:1} \\
 7. & $\mathrm{S(^1D) + CO_2 \rightarrow S + CO_2}$ & Relaxation  & $1.00(-11)$ & $0$ & $0$ & Collision theory\textsuperscript{1} \\
 8. & $\mathrm{S(^1D) + N_2 \rightarrow S + N_2}$ & Relaxation  & $1.00(-11)$ & $0$ & $0$ & Collision theory\textsuperscript{1} \\
 9. & $\mathrm{SO(^1\Delta) + CO_2 \rightarrow SO + CO_2}$ & Relaxation  & $1.00(-11)$ & $0$ & $0$ & Collision theory\textsuperscript{1} \\
 10. & $\mathrm{SO(^1\Delta) + N_2 \rightarrow SO + N_2}$ & Relaxation  & $1.00(-11)$ & $0$ & $0$ & Collision theory\textsuperscript{1} \\
 & & & & & & \\
 \hline
 & & & \multicolumn{2}{c|}{Branching ratio} & \multicolumn{2}{c|}{Reference/Source} \\
 \hline
 11. & $\mathrm{CS_2 + h\nu \rightarrow CS + S(^1D)}$ & Photodissociation & \multicolumn{2}{c|}{ } & \multicolumn{2}{c|}{\textsc{PHIDRATES}\textsuperscript{2,3,4}, JPL\textsuperscript{5}} \\
 12. & $\mathrm{OCS + h\nu \rightarrow CO + S(^1D)}$ & Photodissociation & \multicolumn{2}{c|}{ } & \multicolumn{2}{c|}{\textsc{PHIDRATES}\textsuperscript{2,3,4}, JPL\textsuperscript{5}} \\
 13. & $\mathrm{SO + h\nu \rightarrow S + O}$ & Photodissociation & \multicolumn{2}{c|}{0.75} & \multicolumn{2}{c|}{\textsc{PHIDRATES}\textsuperscript{2,3,4,6}} \\
 & $\mathrm{SO + h\nu \rightarrow S(^1D) + O}$ & Photodissociation & \multicolumn{2}{c|}{0.25} & \multicolumn{2}{c|}{\textsc{PHIDRATES}\textsuperscript{2,3,4,6}} \\
14. & $\mathrm{SO_2 + h\nu \rightarrow SO + O}$ & Photodissociation & \multicolumn{2}{c|}{0.75} & \multicolumn{2}{c|}{\textsc{PHIDRATES}\textsuperscript{2,3,4,7}} \\
 & $\mathrm{SO_2 + h\nu \rightarrow SO(^1\Delta) + O}$ & Photodissociation & \multicolumn{2}{c|}{0.25} & \multicolumn{2}{c|}{\textsc{PHIDRATES}\textsuperscript{2,3,4,7}} \\
15. & $\mathrm{S_2O + h\nu \rightarrow SO + S}$ & Photodissociation & \multicolumn{2}{c|}{0.75} & \multicolumn{2}{c|}{\textsc{PHIDRATES}\textsuperscript{2,3,4,8}}\\
 & $\mathrm{S_2O + h\nu \rightarrow SO(^1\Delta) + S}$ & Photodissociation & \multicolumn{2}{c|}{0.25} & \multicolumn{2}{c|}{\textsc{PHIDRATES}\textsuperscript{2,3,4,8}}\\
16. & $\mathrm{SO(^1\Delta) + h\nu \rightarrow S + O}$ & Photodissociation & \multicolumn{2}{c|}{0.75} & \multicolumn{2}{c|}{\textsc{PHIDRATES}\textsuperscript{2,3,4,9}} \\
17. & $\mathrm{SO(^1\Delta) + h\nu \rightarrow S(^1D) + O}$ & Photodissociation & \multicolumn{2}{c|}{0.25} & \multicolumn{2}{c|}{\textsc{PHIDRATES}\textsuperscript{2,3,4,10}} \\
\hline
 \multicolumn{7}{l}{
 \parbox{\textwidth}{\small\raggedright
 \vspace{0.3cm}
 \textbf{Note.}\\
 \textsuperscript{1} From collision theory, the pre-exponential factor for radical–radical or radical–neutral reactions assuming a barrierless potential is estimated as $\sigma \langle v \rangle = \pi d^2 \times \sqrt{8k_BT/\pi m} \sim 10^{-11}$ cm$^{3}$ molecule$^{-1}$ s$^{-1}$. \\
 \textsuperscript{2} \cite{Huebner_Carpenter_1979}, \textsuperscript{3} \cite{Huebner_1992}, \textsuperscript{4} \cite{huebner2015photoionization}, \textsuperscript{5} \cite{burkholder2019evaluation19}\\
 \textsuperscript{6} Bond dissociation threshold of SO: Ground state channel $\sim$ 2295 \AA, Excited state channel $\sim$ 1894 \AA \, \citep{Darwent1970,di2025hot}\\
 \textsuperscript{7} Bond dissociation threshold of \ce{SO2}: Ground state channel $\sim$ 2173 \AA, Excited state channel $\sim$ 1909 \AA \, \citep{Gope2017}\\
 \textsuperscript{8} Bond dissociation threshold of \ce{S2O}: Ground state channel $\sim$ 3606 \AA, Excited state channel $\sim$ 2934 \AA \, \citep{Han2008,Berkowitz1977,borin1999lowest}\\
 \textsuperscript{9} Photodissociation cross-section is assumed same as that of the channel $\mathrm{SO + h\nu \rightarrow S + O}$\\
 \textsuperscript{10} Photodissociation cross-section is assumed same as that of the channel $\mathrm{SO + h\nu \rightarrow S(^1D) + O}$\\ 
 }
 }
 \end{tabular}
\end{table*}

\subsubsection{Reaction kinetics of \ce{S(^3P)} + \ce{CO2} and \ce{S(^1D)} + \ce{CO2} and new thermochemical Data} \label{sec:kinetic results}

\begin{figure}
\includegraphics[width=\columnwidth]{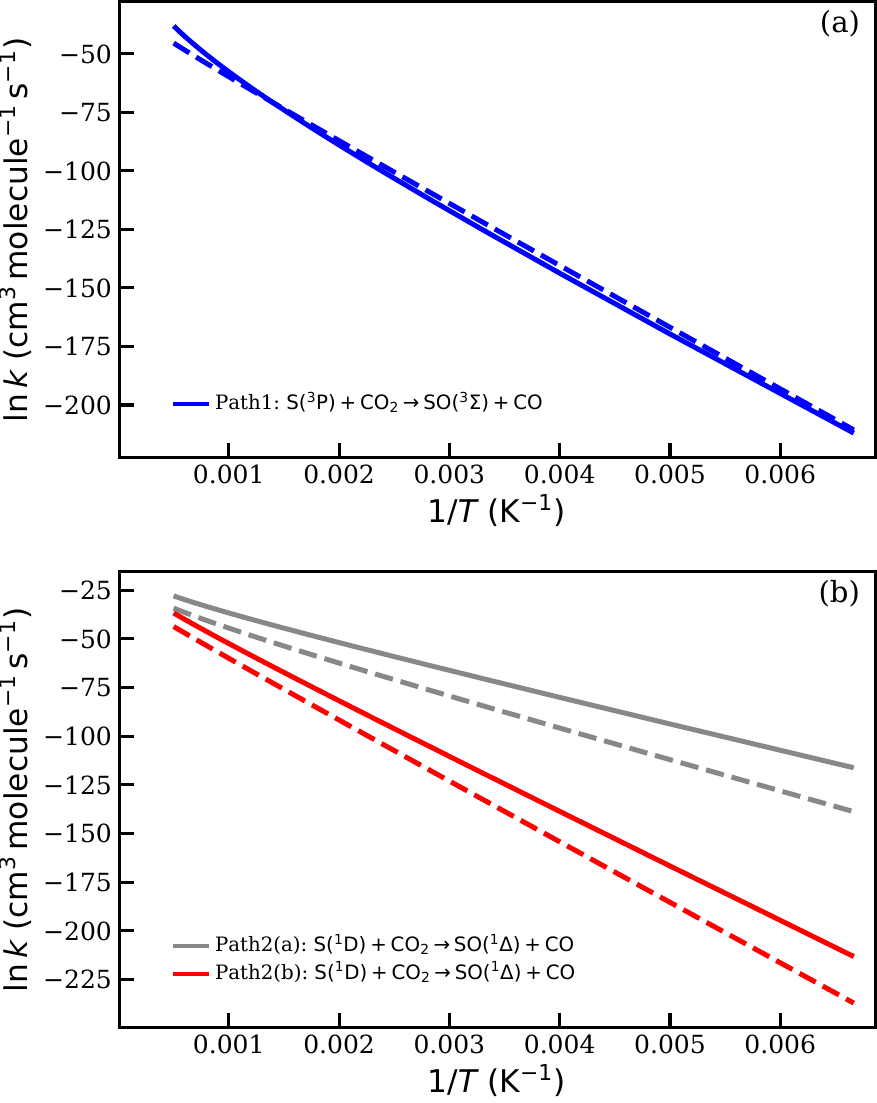}
\caption{Calculated temperature dependence of the rate coefficients over 150 K-2000 K.
\tikz[baseline=-0.5ex]\draw[line width=1pt] (0,0)--(0.6,0); denotes the forward reaction, while
\tikz[baseline=-0.5ex]\draw[dashed,line width=1pt] (0,0)--(0.6,0); depicts the corresponding backward reaction.}
    \label{fig:2}
   \end{figure}

\begin{table*}[ht!]
    \centering
    \caption{NASA 7-term polynomial coefficients for S and SO and their excited states S($^1$D) and $\mathrm{SO}(^1\Delta)$. Calculated values for S and SO molecules are shown with the the original data from \cite{burcat_2005} as benchmark. First and second row correspond to the low and high temperature ranges respectively.}
    \label{tab:nasa7_table}
    \begin{tabular}{|l|rrrrrrr|}
         \hline
         \textbf{Species} & $\mathrm{a_1}$ & $\mathrm{a_2}$ & $\mathrm{a_3}$ & $\mathrm{a_4}$ & $\mathrm{a_5}$ & $\mathrm{a_6}$ & $\mathrm{a_7}$ \\
         \hline
         S (Burcat) (200--1000 K) & 2.3173 & 4.7802e-03 &-1.4208e-05 & 1.5657e-08 &-5.9659e-12 & 3.2507e+04 & 6.0624 \\
         S (Burcat) (1000--6000 K) & 2.8794 &-5.1105e-04 & 2.5381e-07 &-4.4546e-11 & 2.6672e-15 & 3.2501e+04 & 3.9814 \\
         \hline
          S (this work) (10--1363.47 K) & 2.5 & -1.5266e-15 & 6.3226e-18 & -7.59355e-21 & 2.67359e-24 & 33409.9 & 5.12978 \\
          S (this work) (1363.47--3000 K) & 2.5 & -1.01797e-13 & 8.3758e-17 & -2.93808e-20 & 3.72101e-24 & 33409.9 & 5.12978 \\
         \hline
          S($^1$D) (this work) (10--1363.47 K) & 2.5 & -1.5266e-15 & 6.3226e-18 & -7.59355e-21 & 2.67359e-24 & 38017.9 & 4.03117 \\ 
          S($^1$D) (this work) (1363.47--3000 K) & 2.5 & -1.01797e-13 & 8.3758e-17 & -2.93808e-20 & 3.72101e-24 & 38017.9 & 4.03117 \\
         \hline

         SO (Burcat) (200--1000 K) & 3.6186 & -2.3217e-03 &  1.1646e-05 & -1.4209e-08 & 5.6077e-12 & -4.8062e+02 &  6.3650 \\
         SO (Burcat) (1000--6000 K) & 3.9689 & 3.7730e-04 & 7.6710e-09 & -1.3754e-11 & 1.3714e-15 & -7.2857e+02 & 3.7349 \\
         \hline
         SO (this work) (10--877.148 K) & 3.51729 & -0.0010721 & 6.77542e-06 & -7.79703e-09 & 2.85803e-12 & 330.623 & 6.69922 \\
          SO (this work) (877.148--3000 K) & 3.1173 & 0.00225842 & -1.49624e-06 & 4.4775e-10 & -4.99074e-14 & 342.841 & 8.24617 \\
         \hline
          $\mathrm{SO}(^1\Delta)$ (this work) (10--879.964 K) & 3.51727 & -0.00106791 & 6.72899e-06 & -7.71761e-09 & 2.81931e-12 & 4071.98 & 5.59922 \\
          $\mathrm{SO}(^1\Delta)$ (this work) (879.964--3000 K) & 3.11517 & 0.00225782 & -1.49355e-06 & 4.46351e-10 & -4.96919e-14 & 4084.75 & 7.15814 \\
         \hline
    \end{tabular}    
\end{table*}

The temperature-dependent rate coefficients for both the forward and backward directions of Path~1, \ce{S(^3P)} + \ce{CO2}, and Path~2(a) and Path~2(b), \ce{S(^1D)} + \ce{CO2}, are shown in Figure~\ref{fig:2} for the 150--2000~K temperature range. The results clearly indicate that the formation of sulfur monoxide from \ce{S(^3P)} + \ce{CO2} is slower than that from the reaction of excited \ce{S(^1D)} with \ce{CO2}. This behavior is attributed to the high stability of carbon dioxide in its ground state, which corresponds to the large activation barrier required to form the triplet reaction complex ($^3$RC in Figure~\ref{fig:1}) leading to SO.  

The formation of singlet sulfur monoxide proceeds faster via Path~2(a) than via Path~2(b) (Figure~\ref{fig:1}). The barrierless formation of the reaction complex ($^1$RC) and the smaller single-step barrier in Path~2(a) ($^1$TS$\rm _{2a}$) result in higher rate coefficients compared with Path~2(b). In contrast, Path~2(b) involves two steps involving cyclization associated with substantial energy barriers, yielding rate coefficients about three orders of magnitude lower than those of Path~2(a). For both Path~2(a) and Path~2(b), the reverse reactions remain consistently slower than their forward counterparts across the entire temperature range, since the forward reaction is exothermic.  

Path~1 shows an unusual trend (Figure~\ref{fig:2}). While the reverse reaction dominates at low temperatures due to its lower thermodynamic barrier, the forward rate overtakes at higher temperatures. This crossover arises from entropic contributions that reduce the activation free energy,  
\[
\Delta G^\ddagger = \Delta H^\ddagger - T \Delta S^\ddagger,
\]  
making the forward process entropically favored at low temperatures. However, it has to be noted that the rate coefficient at very low temperature ($T\leq$300 K is highly sensitive to the uncertainty in the calculated $\gamma$ term shown in Equation~\ref {eq:arrhenius_eqn}).

Table~\ref{tab:merged_reactions} summarizes the Modified Arrhenius parameters for the 150--2000~K range for the \ce{S} + \ce{CO2} reactant system in both ground and first excited states. While the excited-state channels may proceed through different intermediates, they all converge to the same products, \ce{SO} + \ce{CO}. Notably, the activation temperature ($\gamma$) for all pathways remain on the order of 10$^4$~K.

\begin{figure}[ht!]
    \centering
    \includegraphics[width=\linewidth]{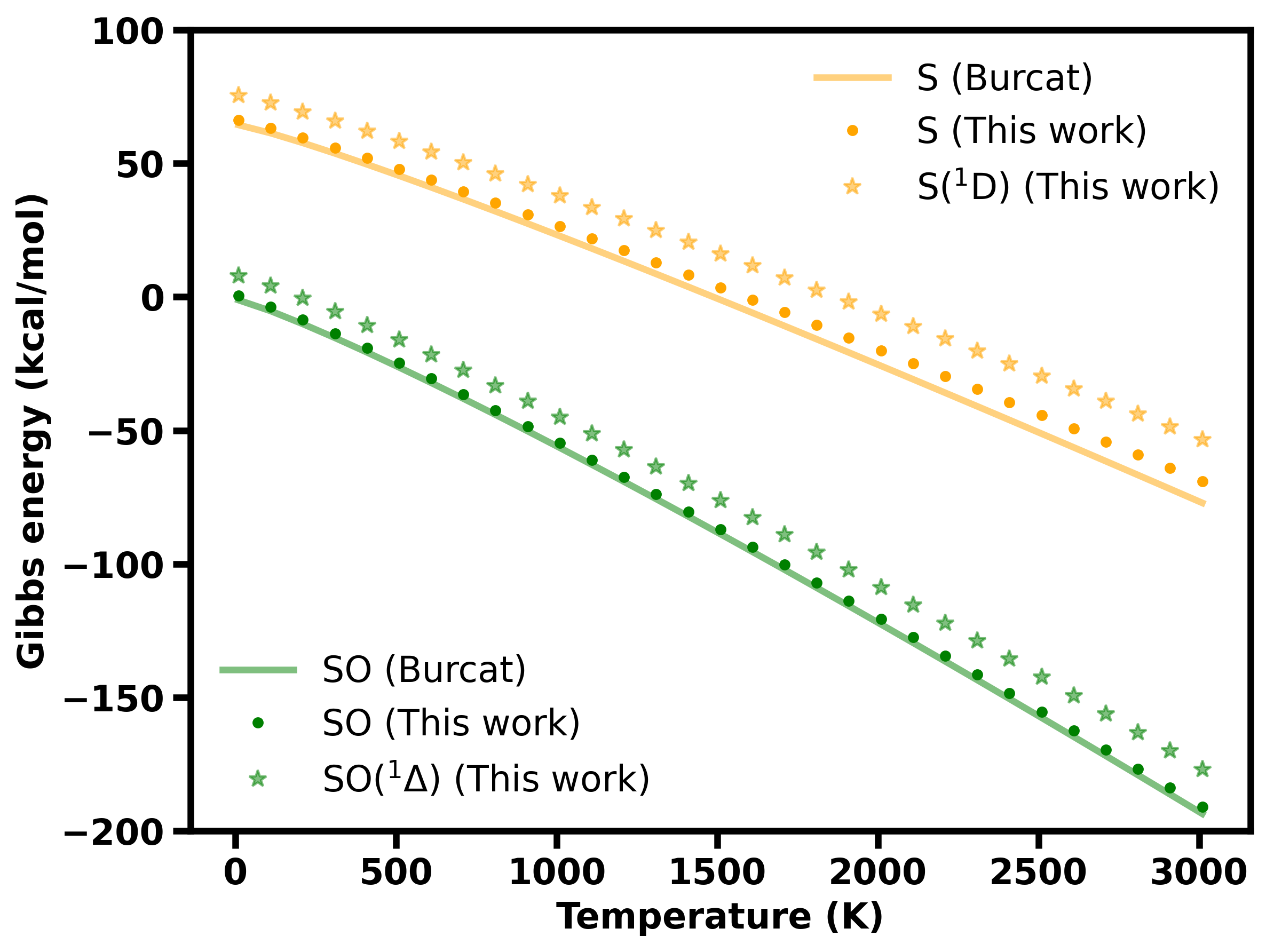}
    \caption{Comparison of Gibbs free energy (kcal/mol) as a function of temperature using NASA polynomial coefficients. Solid lines represent values from the Burcat thermochemical database (ground state), while dotted lines show ground-state values computed with \texttt{Gaussian} and \texttt{Arkane}, as listed in Table~\ref{tab:nasa7_table}. The Gibbs free energies for the corresponding excited states, S($^1$D) and $\mathrm{SO}(^1\Delta)$, computed in this work, are shown with star-shaped markers.}

    \label{fig:Gibbs_S_SO}
\end{figure}

Since the photochemical kinetics model also uses Gibbs free energy to compute equilibrium constants and reverse reaction rates, we provided new thermochemical data for S($^1$D) and $\mathrm{SO}(^1\Delta)$. These thermodynamic parameters were first validated for the ground-state sulfur species (S and SO) using the Burcat thermochemical database, as shown in Figure~\ref{fig:Gibbs_S_SO}, to assess the accuracy and reliability of our computed NASA 7-term polynomial coefficients for S($^1$D) and $\mathrm{SO}(^1\Delta)$, which are listed in Table~\ref{tab:nasa7_table}. 

\begin{figure*}[ht]
    \centering
    \includegraphics[width=\linewidth]{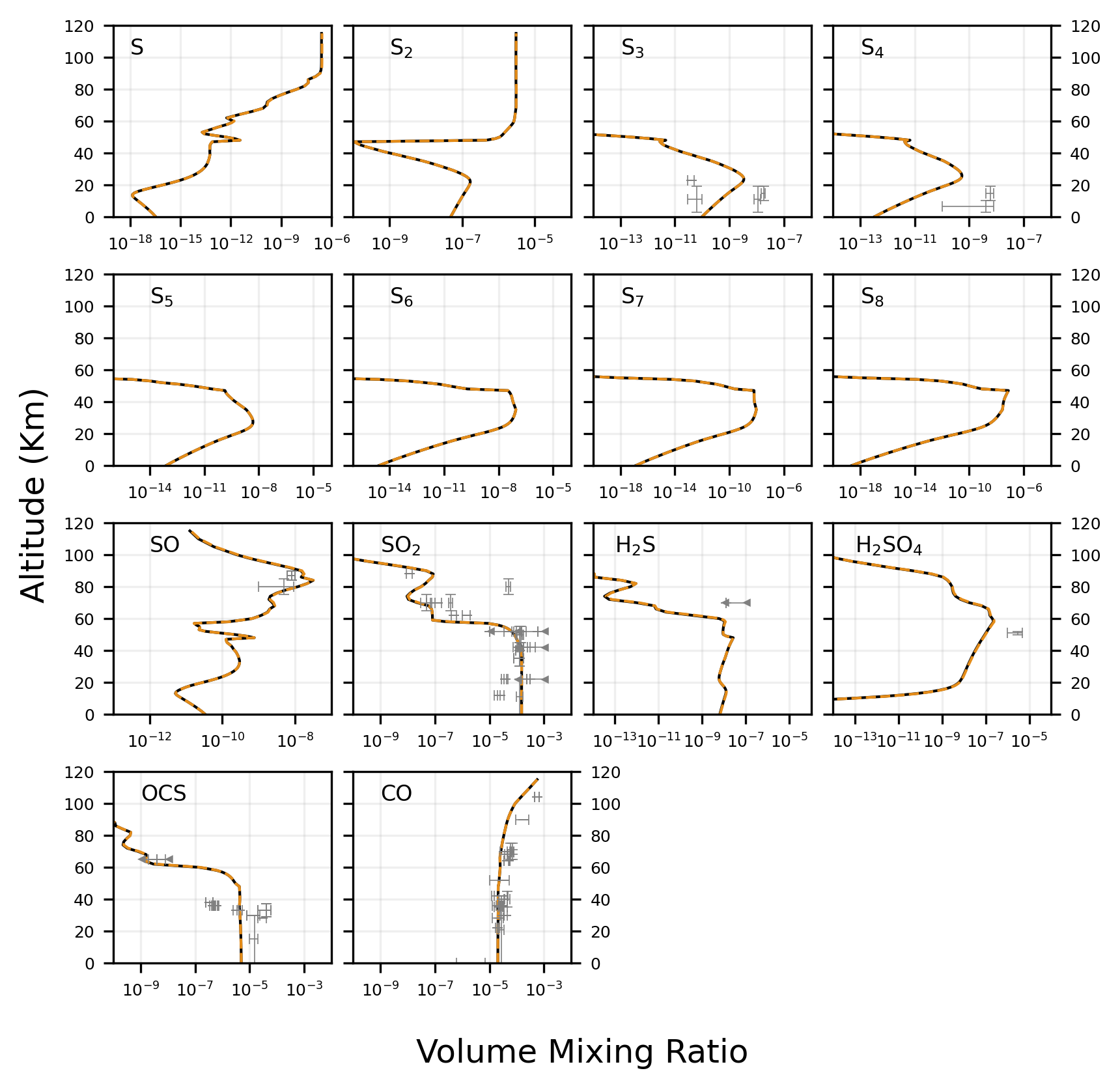}
    \caption{Predicted volume mixing ratios of S$_n$ ($\mathrm{n=1\text{--}8}$), SO, SO$_2$, H$_2$S, H$_2$SO$_4$, OCS, and CO (including both gaseous and condensed S$_2$ and H$_2$SO$_4$) as a function of altitude (km) in the Venusian atmosphere, using the \texttt{ARGO-STAND2020} network (solid black, similar to the cloud‐chemistry model of \cite{Rimmer_2021}) and the \texttt{XODIAC-2025.v1} network (dashed orange). Data from various Venus missions, listed in Table~4 of \cite{Rimmer_2021}, are also shown in gray. Although observational data for polysulfurs S$_2$, S$_5$, S$_6$, S$_7$, and S$_8$ are not available, we include them to compare polysulfur chemistry between the \texttt{ARGO-STAND2020} and \texttt{XODIAC-2025.v1} networks, since S$_3$ and S$_4$ deviate from observed values in both cases.}
    \label{fig:benchmark_venus_vmr_stand2020_xodnet}
\end{figure*}

\subsection{Astrophysical Implications from Photochemical Kinetics Modeling with \texttt{XODIAC}}

\subsubsection{Validation of \texttt{XODIAC} and \texttt{ARGO} for Venusian Sulfur Chemistry with the \texttt{XODIAC-2025} and \texttt{STAND-2020} Networks}

To validate our photochemical model \texttt{XODIAC} \citep{Ghosh_2026} for Venus using the \texttt{XODIAC-2025.v1} chemical network, we benchmarked our Venusian simulations against the \texttt{STAND-2020} network, which was previously used by \citet{Rimmer_2021} with the photochemical–diffusion code \texttt{ARGO}. For this comparison, we incorporated into \texttt{XODIAC} the condensation scheme based on the modified Antoine equations from \citet{Rimmer_2016,Rimmer_2021}, as well as the parameterization of Venus' unknown UV absorber from \citet{KRASNOPOLSKY2012,Rimmer_2021,Greaves2020}.

For benchmarking, we adopted the same initial surface mixing ratios, T–P–K$_{zz}$ profile, and incident solar flux at the top of the atmosphere as \citet{Rimmer_2021}. We find strong agreement between the \texttt{XODIAC-2025.v1} network and \texttt{ARGO}+\texttt{STAND-2020} for all major sulfur-bearing species in the Venusian atmosphere. The model results are shown in Figure~\ref{fig:benchmark_venus_vmr_stand2020_xodnet}, together with observed abundances from various Venus missions plotted with error bars. The observational dataset is compiled from \citet{Arney2014, Bertaux1996, Bezard1990, Bezard2007, Collard1993, Connes1968, Cotton2012, Encrenaz2012, Encrenaz2015, Fegley2014, Gelman1979, Grassi2014, Greaves2020, Hoffman1980, Kras2008, Kras2010, Kras2013, Kras2014, Maiorov2005, Marcq2005, Marcq2006, Marcq2008, Marcq2015, Mukhin1982, Mukhin1983, Na1990, Oschlisniok2012, Oyama1979, Oyama1980, Pollack1993, Tsang2008, Wilson1981, Young1972, Zasova1993}, as summarized by \citet{Rimmer_2021}. 

It is evident from Figure~\ref{fig:benchmark_venus_vmr_stand2020_xodnet} that most sulfur-bearing species are well reproduced by both \texttt{XODIAC}+\texttt{XODIAC-2025.v1} and \texttt{ARGO}+\texttt{STAND-2020}, with the exception of \ce{S3} and \ce{S4}.

\subsubsection{Does the Extended \texttt{XODIAC} Chemical Network \texttt{XODIAC-2025.v2}, with Excited \ce{S(^1D)} and \ce{SO(^1\Delta)}, Produce Observable Differences in Venusian Sulfur Chemistry?}

The extended \texttt{XODIAC-2025} network \texttt{XODIAC-2025.v2} includes our calculated $\mathrm{S\!-\!CO_2}$ reactions, an $\mathrm{S\!-\!CO}$ reaction, relaxation pathways for $\mathrm{S(^1D)}$ and $\mathrm{SO(^1\Delta)}$, and photochemical channels involving \ce{CS2}, \ce{OCS}, \ce{SO}, $\mathrm{SO(^1\Delta)}$, \ce{SO2}, and \ce{S2O}. We now investigate whether including sulfur reaction kinetics in both ground and excited states, together with the associated reactions listed in Table~\ref{tab:merged_reactions}, affects the atmospheric chemistry of Venus when modeled using \texttt{XODIAC} with the \texttt{XODIAC-2025.v2} chemical network.

\begin{figure*}[ht]
    \centering
    \includegraphics[width=\linewidth]{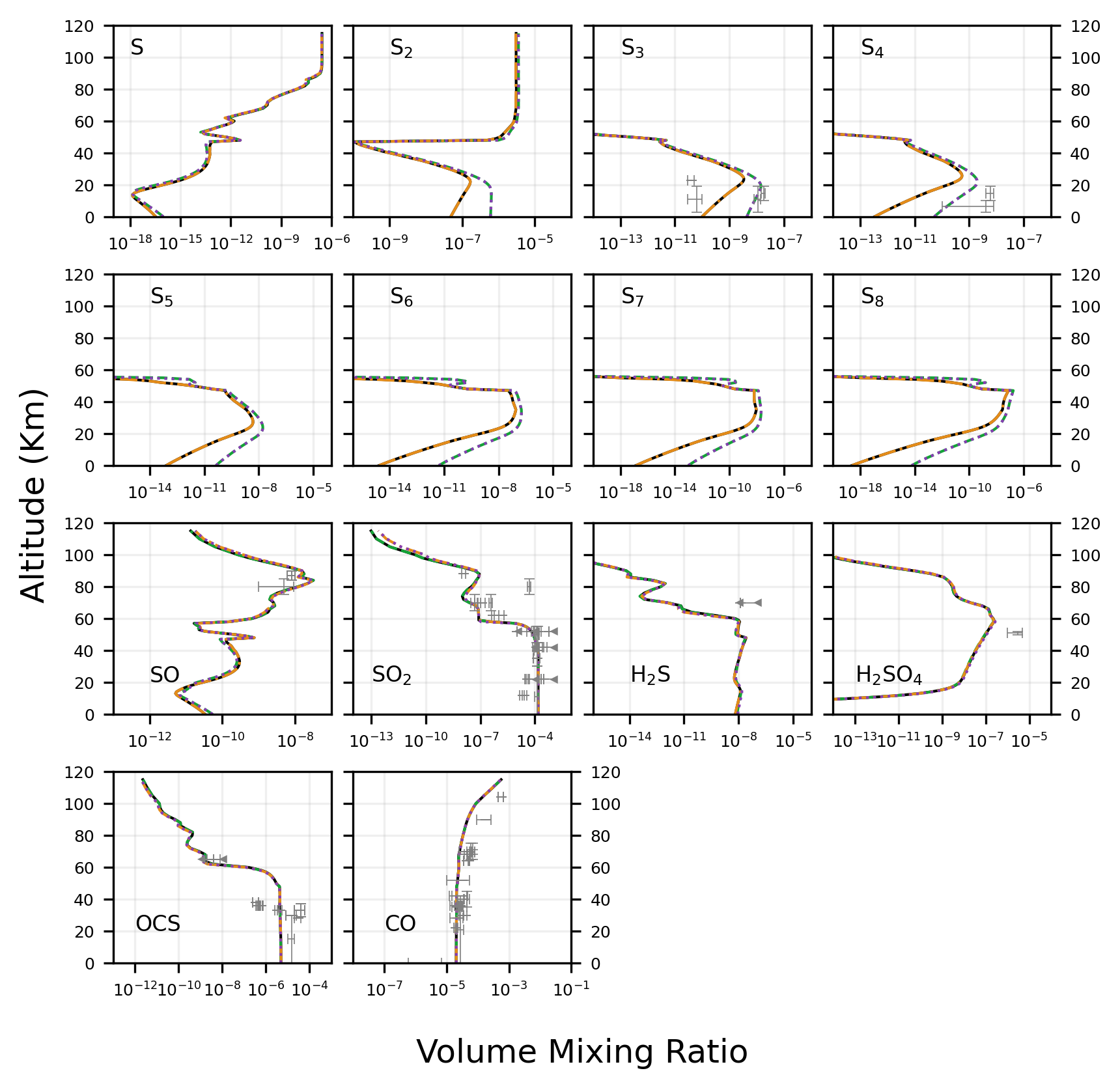}
    \caption{Predicted volume mixing ratios of S$_n$ ($\mathrm{n=1\text{--}8}$), SO, SO$_2$, H$_2$S, H$_2$SO$_4$, OCS, and CO (including both gaseous and condensed S$_2$ and H$_2$SO$_4$) as a function of altitude (km) in the Venusian atmosphere, using the \texttt{XODIAC-2025.v1} network (solid black; M0), the \texttt{XODIAC-2025.v1} network with a 1~ppm surface mixing ratio of atomic sulfur (dashed green; M1), the \texttt{XODIAC-2025.v2} network (dashed-dotted orange; M2), and the \texttt{XODIAC-2025.v2} network with a 1~ppm surface mixing ratio of atomic sulfur (dotted purple; M3). Data from various Venus missions, listed in Table~4 of \cite{Rimmer_2021}, are also shown in gray. Although observational data for polysulfurs S$_2$, S$_5$, S$_6$, S$_7$, and S$_8$ are not available, we include them to compare polysulfur chemistry across the networks. We do not show $\mathrm{S(^1D)}$ and $\mathrm{SO(^1\Delta)}$, as their mixing ratios reach a maximum of only $\sim 10^{-15}$, which lies far below observational detectability.}

    \label{fig:venus_vmr}
\end{figure*}

We compare the modeled abundances of major sulfur-bearing species and CO using both the base \texttt{XODIAC-2025.v1} network and the extended \texttt{XODIAC-2025.v2} network, as shown in Figure~\ref{fig:venus_vmr}. Observational data from various Venus missions are also plotted with error bars. It is evident from Figure~\ref{fig:venus_vmr} that most sulfur-bearing species are reproduced within observational uncertainties by the \texttt{XODIAC-2025.v2} network, similar to the base network, with the exceptions of \ce{S3} and \ce{S4}, which remain underproduced in both cases. This indicates that the overall Venusian sulfur chemistry remains largely unchanged even after adding the new reactions. A closer examination shows that a few photochemically active species above roughly 60 km, such as SO, \ce{SO2}, and \ce{OCS}, show small shifts in their mixing ratios, none of which alter the agreement with observed abundances. These minor variations arise primarily from differences in the rates of several photochemical pathways. A detailed diagram of the dominant reaction pathways and their corresponding rates for SO, SO$_2$, and OCS is provided in Figure~\ref{fig:soso2h2s_m0m2} in the Appendix.



\begin{figure*}[ht]
    \centering
    \includegraphics[width=\linewidth]{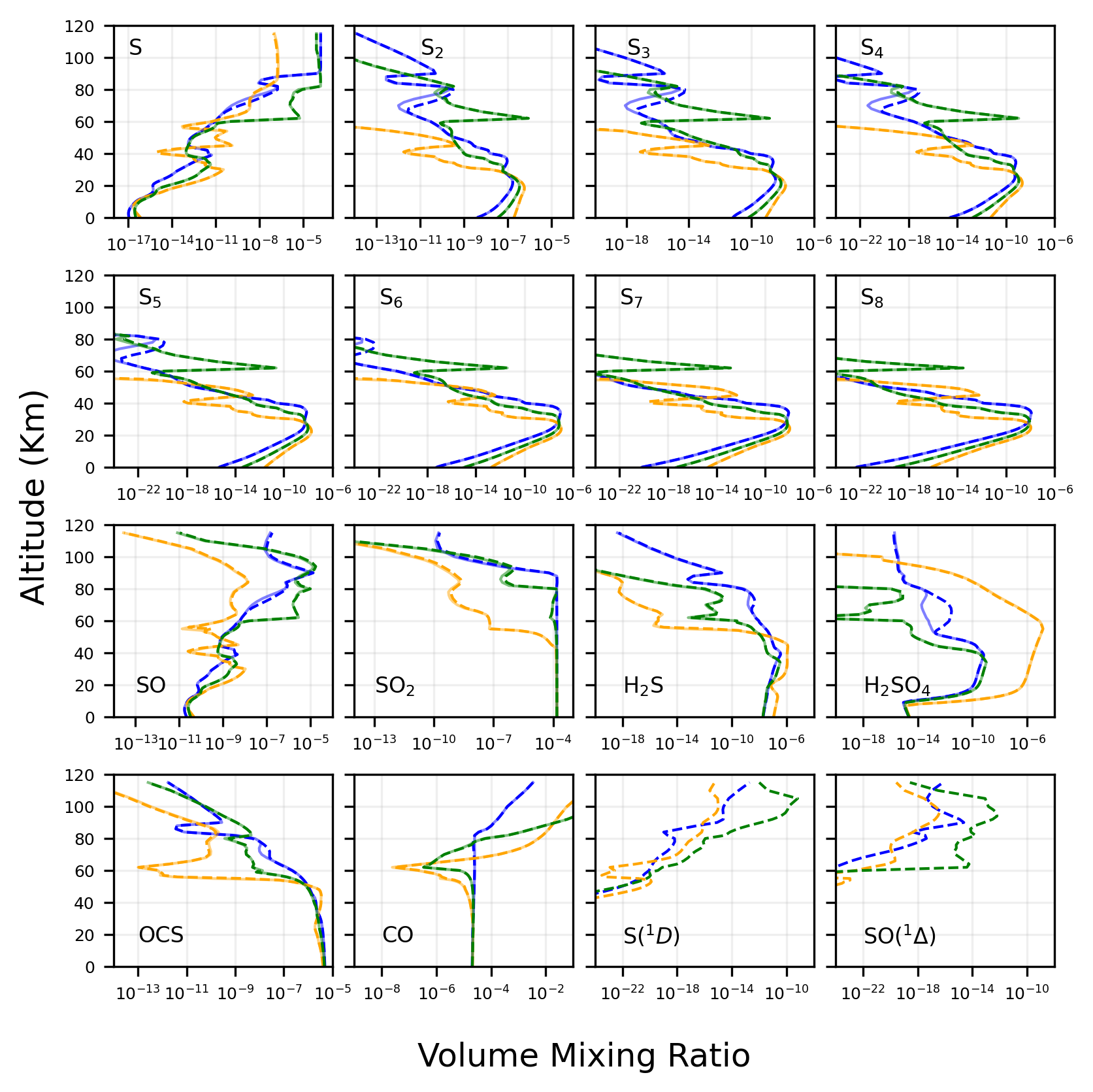}
  \caption{Predicted volume mixing ratios of S$_n$ ($\mathrm{n=1\text{--}8}$), SO, SO$_2$, H$_2$S, H$_2$SO$_4$ (gas + condensed), OCS, CO, S($^1$D) and SO($^1\Delta$) as a function of altitude (km) in the exo-Venus atmosphere, using three models: M4 (blue; exo-Venus atmospheric profile from \cite{KRASNOPOLSKY2007} with a 505.6~K isotherm), M5 (orange; 1000$\times$ solar flux at the top of the atmosphere), and M6 (green; combination of M4 and M5). Solid lines represent results from the \texttt{XODIAC-2025.v1} network, and dashed lines represent those obtained with the \texttt{XODIAC-2025.v2} network (M4, M5, M6).}

    \label{fig:exovenus_vmr}
\end{figure*}

\begin{figure*}[ht]
    \centering
    \includegraphics[width=\linewidth]{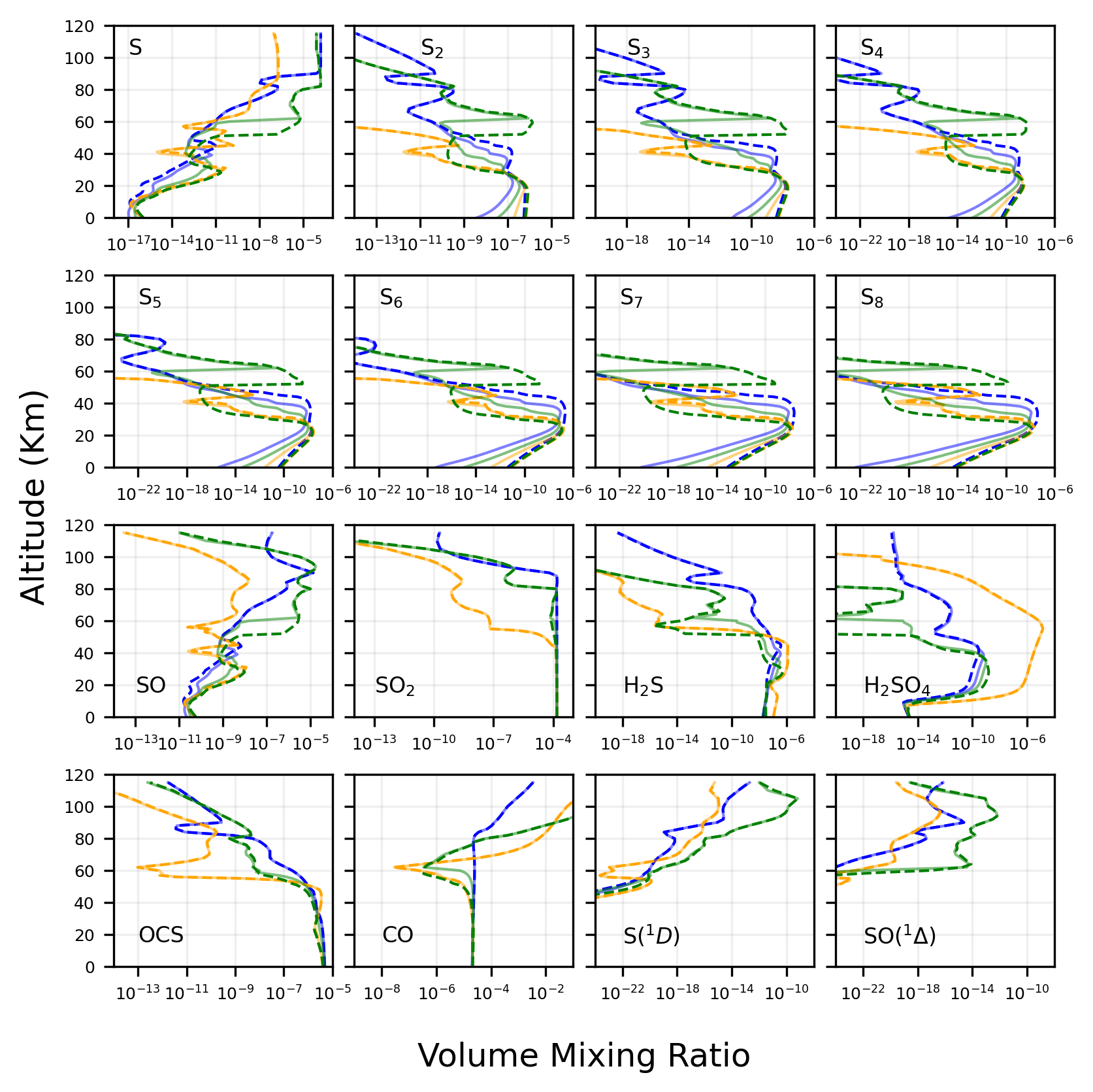}
  \caption{Predicted volume mixing ratios of S$_n$ ($\mathrm{n=1\text{--}8}$), SO, SO$_2$, H$_2$S, H$_2$SO$_4$ (gas + condensed), OCS, CO, S($^1$D) and SO($^1\Delta$) as a function of altitude (km) in the exo-Venus atmosphere, using three model comparisons: M4 vs M7 (blue; exo-Venus atmospheric profile from \cite{KRASNOPOLSKY2007} with a 505.6~K isotherm), M5 vs M8 (orange; 1000$\times$ solar flux at the top of the atmosphere), and M6 vs M9 (green; combination of M4 and M5 or M7 and M8). Solid lines represent results from the \texttt{XODIAC-2025.v2} network with S = 0 (M4, M5, M6), and dashed lines represent those obtained with the \texttt{XODIAC-2025.v2} network with S = 1 ppm (M7, M8, M9).}

    \label{fig:exovenus_vmr_1ppm}
\end{figure*}

\subsubsection{Accounting for Missing Sulfur Processes on Venus} \label{sec:nonzero_sulfur}

Even with the extended \texttt{XODIAC-2025.v2} network, the modeled abundances of \ce{S3} and \ce{S4} remain significantly lower than the observed values. This discrepancy motivates a closer examination of the missing sulfur processes in both the \texttt{XODIAC-2025.v1} and \texttt{XODIAC-2025.v2} networks. In Figure~\ref{fig:venus_vmr}, we therefore include cases from both networks in which a non-zero lower boundary atomic sulfur abundance of 1~ppm is imposed to account for these missing processes indirectly. This boundary condition serves as a diagnostic of the net flux of reduced sulfur from the deep atmosphere or surface that is not explicitly represented in our model. The introduction of sulfur atoms implies that, if approximately one percent of sulfur-bearing species below the cloud layer is chemically activated and capable of releasing atomic sulfur, the observed abundances of \ce{S3} and \ce{S4} can be reproduced.

There are potential sources of activated sulfur that would behave similarly to the imposed atomic sulfur. One possibility is a simple mechanism in which sulfur is liberated from condensed phases into the gas phase. Perhaps atomic sulfur is being released from Venus's surface. Less likely, the sulfur could be coming from the cloud droplets as they evaporate at the bottom of the clouds. This seems less likely because the bulk of the droplet chemistry is believed to be more oxidizing than the surrounding atmosphere \citep{KRASNOPOLSKY2012,Jiang2024}. The surface seems more likely, though the exchange of atoms, or really, electrons, between the lower atmosphere and the surface is uncertain: it may well be the atmosphere and surface is not in redox equilibrium \citep{Bierson2020}. 

Another, more general possibility is that the reactive sulfur is sequestered in another chemical form. The atmosphere between 48~km and the surface may host a species capable of releasing a sulfur atom under perturbation, such as \ce{S2O}, although only if it is far more abundant than current models predict. It is also possible that, at the higher temperatures below the clouds, some sulfur chains become strained or develop terminal sulfur atoms bound by a single bond, analogous to loose threads. Although such fragments would constitute only a minor fraction of the sulfur inventory, they may nonetheless be sufficient to supply the sulfur required for this chemistry.

The \ce{S3} and \ce{S4} observations are highly uncertain, because it is difficult to infer specific species from such broad observed features. If the observations are accurate, then our predictions favor the presence of some lower-boundary source of active sulfur. This hypothesis can be tested either by lower-atmosphere observations of the sulfur cycle, for example, by \texttt{DAVINCI} \citep{Garvin2022}, in tandem with further laboratory and theoretical exploration into the behavior of elemental sulfur under high temperatures and thick \ce{CO2} gas, or with minerals in contact with supercritical \ce{CO2}. In other words, exploring sulfur chemistry under simulated extreme conditions of Venus's lower atmosphere.

Motivated by this possibility, we performed simulations using both the \texttt{XODIAC-2025.v1} and \texttt{XODIAC-2025.v2} chemical networks that include a non-zero lower-boundary atomic sulfur abundance. These models yield a substantially improved agreement with the observed polysulfur abundances (\ce{S3} and \ce{S4}). In particular, the resulting near-surface mixing ratios fall within the observational error bars from Venus mission data, while producing only minor changes in the abundances of other sulfur-bearing species. We further explored initial atomic sulfur abundances in the range $10^{-8}$ to $10^{-5}$; however, in all such cases the resulting \ce{S3} and \ce{S4} vertical profiles deviated from the observational constraints, including their associated uncertainties. Notably, the abundances of polysulfurs increase by approximately a factor of two with each additional sulfur atom (\ce{S2}, \ce{S3}, and \ce{S4}), highlighting the nonlinear nature of polysulfur chemistry driven by numerous interconversion reactions operating throughout the lower and middle atmosphere.

In regions where thermochemistry dominates, namely at pressures above 10~bar (corresponding to altitudes of approximately 30~km and below), the major reactions controlling the chemistry differ between models M0 and M1, as well as between M2 and M3. All polysulfur species ($\mathrm{S_n}$) are efficiently interconverted, and the dominant reactions responsible for these processes are listed in Table~\ref{tab:polys_reactions_1} for \ce{S2}, \ce{S3}, and \ce{S4}. At higher altitudes (around 54~km), deep within the Venusian cloud deck, condensation of sulfur species such as S$_2$ proceeds rapidly. The major conversion pathways operating at these altitudes are summarized in Table~\ref{tab:reactions_mid_altitude}, where $g$ and $l$ denote gaseous and condensed (liquid) phases, respectively.

For each sulfur species, the net abundance is determined by the balance between its production and loss rates. These rates can differ by several orders of magnitude, even for the same dominant reaction across different models. For example, reaction~1 in Table~\ref{tab:polys_reactions_1} governs both the formation and destruction of S$_2$ in models M0 and M1, contributing approximately 99.8\% of the total rate in both cases, yet the absolute reaction rates differ by nearly two orders of magnitude. Such discrepancies can arise even at the same altitude because the absolute rate of S$_2$ formation or destruction depends on the mixing ratios of all reactants involved, whereas the percentage contribution depends on the sum of all S$_2$-related reaction rates within the network. 


\begin{table}[ht!]
\centering
\caption{Sulfur reaction pathways at lower altitudes ($\lesssim$ 30 km).}
\label{tab:polys_reactions_1}
\begin{tabular}{|l|l|l|}
\hline
\# & Reaction & Reaction Type \\
\hline
1. & $\mathrm{2\,S_2 + M \leftrightharpoons S_4 + M}$ & Three-body \\
2. & $\mathrm{S_2 + S_6 + M \leftrightharpoons S_8 + M}$ & Three-body \\
3. & $\mathrm{S_2 + S_3 + M \leftrightharpoons S_5 + M}$ & Three-body \\
4. & $\mathrm{S + S_2 + M \rightarrow S_3 + M}$ & Three-body \\
5. & $\mathrm{S_3 + S_6 \leftrightharpoons S_2 + S_7}$ & Two-body \\
6. & $\mathrm{S_3 + S_7 \leftrightharpoons S_2 + S_8}$ & Two-body \\
7. & $\mathrm{S_3 + S_7 \leftrightharpoons 2\,S_5}$ & Two-body \\
8. & $\mathrm{S_3 + S_7 \leftrightharpoons S_4 + S_6}$ & Two-body \\
9. & $\mathrm{S_4 + S_6 \leftrightharpoons S_2 +S_8}$ & Two-body \\
10. & $\mathrm{S_3 + S_8 \leftrightharpoons S_4 + S_7}$ & Two-body \\


11. & $\mathrm{S_4 \xrightarrow{h\nu} 2\,S_2}$ & Photochemical \\
12. & $\mathrm{S_4 \xrightarrow{h\nu} S_3 + S}$ & Photochemical \\


\hline

\end{tabular}
\end{table}

\begin{table}[ht]
\centering
\caption{Sulfur reaction pathways at mid-altitudes ($\sim$54 km).}
\label{tab:reactions_mid_altitude}
\begin{tabular}{|l|l|l|}
\hline
\# & Reaction & Reaction Type \\
\hline
1. & $\mathrm{S_2} (g)$ $\rightarrow$ $\mathrm{S_2} (l)$ & Condensation \\
2. & $\mathrm{S + SNO \rightarrow S_2 + NO}$ & Two-body \\
3. & $\mathrm{S + OCS \rightarrow CO + S_2}$ & Two-body \\
4. & $\mathrm{S_3 \xrightarrow{h\nu} S_2 + S}$ & Photochemical \\
5. & $\mathrm{S_4 \xrightarrow{h\nu} S_3 + S}$ & Photochemical \\
6. & $\mathrm{S + S_8 \rightarrow S_3 + S_6}$ & Two-body \\
7. & $\mathrm{S + S_7 \rightarrow S_3 + S_5}$ & Two-body \\
8. & $\mathrm{2\,S_5 \rightarrow S_3 + S_7}$ & Two-body \\
9. & $\mathrm{S_4 \xrightarrow{h\nu} 2\,S_2}$ & Photochemical \\
10. & $\mathrm{S_4 \xrightarrow{h\nu} S_3 + S}$ & Photochemical \\
11. & $\mathrm{S + S_8 \rightarrow S_4 + S_5}$ & Two-body \\
12. & $\mathrm{S + S_7 \rightarrow 2\,S_4}$ & Two-body \\
13. & $\mathrm{S_2 + S_2 + M \rightarrow S_4 + M}$ & Three-body \\
\hline

\end{tabular}
\end{table}

Notably, the enhanced abundance of sulfur atoms in the deep atmosphere (up to 30~km) approximately doubles the abundance of S$_2$ in models M1 and M3 relative to models M0 and M2, with even larger increases observed for the higher polysulfurs \ce{S3} and \ce{S4}.

These changes in abundance arise primarily from differences in the dominant reaction rates between models M0 and M1, and between models M2 and M3. A higher initial sulfur mixing ratio modifies the early-time production and loss rates of several species, including \ce{SO}, \ce{CO}, and other sulfur-bearing molecules. As the system evolves, the elevated sulfur abundance in models M1 and M3 leads to modest deviations from the corresponding abundances in models M0 and M2. These small differences propagate through the network of interconversion reactions listed in Table~\ref{tab:polys_reactions_1}, ultimately influencing the abundances of the polysulfurs. The dominant reaction channels and their corresponding rates for S$_2$, S$_3$, and S$_4$ are shown in Figures~\ref{fig:s2s3s4_m0m1} and \ref{fig:s2s3s4_m1m2} for the M0 versus M1 and M1 versus M2 comparisons, respectively, in the Appendix.

Overall, imposing a 1~ppm lower-boundary sulfur mixing ratio produces enhancements of one to two orders of magnitude in the modeled S$_3$ and S$_4$ abundances, bringing them substantially closer to the observational constraints. This strong sensitivity indicates that the observed polysulfur chemistry on Venus cannot be reproduced without either an additional source of reduced sulfur from the deep atmosphere or surface, or more efficient formation pathways for S$_3$ and S$_4$ than those currently included in the \texttt{XODIAC-2025.v1} and \texttt{XODIAC-2025.v2} networks. Notably, the updates introduced from \texttt{XODIAC-2025.v1} to \texttt{XODIAC-2025.v2} do not significantly alter the resulting S$_3$ and S$_4$ mixing-ratio profiles. In other words, the missing sulfur problem likely reflects an unmodeled deep-atmosphere sulfur flux, missing reaction channels in the chemical network, or an underestimation of the efficiencies of the known polysulfur-forming pathways.

\subsubsection{Does the Extended \texttt{XODIAC} Chemical Network \texttt{XODIAC-2025.v2}, Including Excited \ce{S(^1D)} and \ce{SO(^1\Delta)}, Produce Observable Differences in Sulfur Chemistry for Exo-Venus Analogs?}

Our models M4--M6 represent exo-Venus analog atmospheres with varying temperature--pressure ($T$--$P$) conditions and stellar flux. In this section, we discuss the implications of incorporating the sulfur reaction kinetics from Table~\ref{tab:merged_reactions} into these exo-Venus analogs.

Comparisons between \texttt{XODIAC-2025.v1} (base) and \texttt{XODIAC-2025.v2} reveal only modest differences in the abundances of major sulfur-bearing species (Figure~\ref{fig:exovenus_vmr}). This limited impact arises because (i) the added reactions remain largely inefficient under the considered exo-Venus conditions, and (ii) the lifetimes of excited-state species are significantly shorter than those of their ground-state counterparts, allowing them to relax efficiently via collisions with \ce{CO2} and \ce{N2}:
\begin{equation}\label{eq:co2_relax}
    \mathrm{S(^1D) + CO_2 \rightarrow S + CO_2}
\end{equation}
\begin{equation}\label{eq:n2_relax}
    \mathrm{S(^1D) + N_2 \rightarrow S + N_2}
\end{equation}

An increase in the probability of excited-state channels or in their lifetimes would be expected to produce noticeable changes in the mixing ratios of major sulfur species. The minor differences above $\sim$50~km arise primarily from variations in photochemical reaction rates (see Appendix Figures~\ref{fig:ssoso2_m4}--\ref{fig:ssoso2_m6}), which reflect the inclusion of both ground- and excited-state photoreactions (Table~\ref{tab:merged_reactions}). Among models M4, M5, and M6, model M4 shows the largest changes in sulfur species mixing ratios in Figure~\ref{fig:exovenus_vmr} when compared to the corresponding case using the \texttt{XODIAC-2025.v1} network. The reactions from the \texttt{XODIAC-2025.v2} network contributing to these differences are:
\begin{equation}\label{eq:so--s_o}
    \mathrm{SO \xrightarrow{h\nu} S + O}
\end{equation}
\begin{equation}\label{eq:so2--so_o}
    \mathrm{SO_2 \xrightarrow{h\nu} SO + O}
\end{equation}
\begin{equation}\label{eq:ocs--s_co}
    \mathrm{OCS \xrightarrow{h\nu} S + CO}
\end{equation}
\begin{equation}\label{eq:ocs--s1d_co}
    \mathrm{OCS \xrightarrow{h\nu} S(^1D) + CO}
\end{equation}

The primary species driving changes in sulfur chemistry are S, SO, \ce{SO2}, and \ce{S2O}, which propagate to other species through both thermochemical and photochemical reactions in the network. As in Venus, enhanced production of atomic sulfur, through photochemistry or relaxation of excited-state sulfur, modifies reaction rates across sulfur-bearing pathways and reshapes the vertical distribution of sulfur. This, in turn, regulates polysulfur chemistry. Once initiated, polysulfur formation proceeds through a cascade of reactions that produce higher-order allotropes, ultimately affecting species up to \ce{S8}. For species such as SO and \ce{SO2}, the dominant formation and loss pathways involve heterogeneous chemistry that exchanges sulfur with cloud droplets (see \citet{Rimmer_2021}).

The presence of excited-state channels in the \texttt{XODIAC-2025.v2} network, along with their bond dissociation cutoffs, redistributes incoming stellar energy into otherwise inaccessible pathways. Compared to \texttt{XODIAC-2025.v1}, the \texttt{XODIAC-2025.v2} network reduces the effective photodissociation cross sections of SO, \ce{SO2}, and \ce{S2O} due to branching between ground- and excited-state channels, leading to lower photodissociation rates. Even under 1000$\times$ solar irradiation (M5 and M6), excited-state photochemistry remains relatively inefficient under Venus-like initial conditions, producing only subtle changes in molecular abundances. Although some atmospheric layers may exhibit enhanced contributions from excited-state photochemistry, these effects are unlikely to significantly alter the abundances of other species.

In model M5, the base \texttt{XODIAC-2025.v1} network produces SO additionally through:
\begin{equation}
    \mathrm{S + HO_2 \rightarrow SO + OH}
\end{equation}
whereas the \texttt{XODIAC-2025.v2} network includes additional pathways:
\begin{equation}
    \mathrm{S + SNO \rightarrow S_2 + NO}
\end{equation}
\begin{equation}
    \mathrm{HS + O \rightarrow H + SO}
\end{equation}
as well as reaction~\ref{eq:co2_relax} in M5, and
\begin{equation}
    \mathrm{O + S_2 \rightarrow SO + S}
\end{equation}
in M6. The precursors \ce{HO2} and SNO form via:
\begin{equation}
    \mathrm{O_2 + H + M \rightarrow HO_2 + M}
\end{equation}
\begin{equation}
    \mathrm{S + NO + M \rightarrow SNO + M}
\end{equation}

The reaction pathways for S, SO, and \ce{SO2} are illustrated in Figures~\ref{fig:ssoso2_m4}--\ref{fig:ssoso2_m6}.

In model M6, \ce{S(^1D)} and \ce{SO(^1\Delta)} reach peak mixing ratios of order $10^{-9}$ and $10^{-11}$, respectively, driven by the combined effects of intense 1000$\times$ solar irradiation and the 505.6~K isothermal profile. Photochemistry continuously produces these excited species, which subsequently relax to their ground states, thereby increasing the abundances of S and SO and enhancing their associated reaction pathways. Notably, unlike the Venusian models, reactions involving condensed \ce{SO2} and \ce{H2SO4} are not dominant in M4 or M6 because the constant isotherm above $\sim$30~km prevents efficient operation of:
\begin{equation}
    \mathrm{SO_2 + H_2SO_4(\ell) \leftrightharpoons SO_2(\ell) + H_2SO_4(\ell)}
\end{equation}

These results highlight the importance of a high-altitude isotherm and intense stellar flux in controlling sulfur photochemistry in exo-Venus atmospheres.


\subsubsection{Does a Missing Sulfur Scenario Similar to Venus Produce Observable Differences in Sulfur Chemistry for Exo-Venus Analogs?}

We impose a 1~ppm atomic sulfur abundance in our exo-Venus cases, analogous to the Venusian atmosphere, using the \texttt{XODIAC-2025.v2} network (M7--M9) to assess its impact on sulfur chemistry (see Section~\ref{sec:photomodeling_section}) and to identify the role of reduced sulfur from the deep atmosphere or surface that is not explicitly represented in our model.

As shown in Figure~\ref{fig:exovenus_vmr_1ppm}, the introduction of 1~ppm atomic sulfur substantially alters the vertical distribution of polysulfur species in the lower and mid-atmosphere, as well as other sulfur species (such as \ce{H2S}, \ce{H2SO4}, and SO) and CO in the mid-atmosphere. The comparison M9 vs M6 shows the largest variation across the atmosphere, relative to M7 vs M4 and M8 vs M5. Near the surface, polysulfur species increase by approximately one order of magnitude in M7 compared to M4, followed by progressively smaller changes in M9 vs M6 and M8 vs M5. In contrast to the Venusian models, the exo-Venus models show stronger variations in polysulfur abundances near the bottom of the atmosphere. Due to the enhanced availability of atomic sulfur, the following reactions act as sulfur sinks:
\begin{equation}
    \mathrm{S + S_2 + M \leftrightharpoons S_3 + M}
\end{equation}
\begin{equation}
    \mathrm{S + SO_2 \leftrightharpoons 2\,SO}
\end{equation}
\begin{equation}\label{eq:s_ocs--co_s2}
    \mathrm{S + OCS \leftrightharpoons CO + S_2}
\end{equation}
\begin{equation}
    \mathrm{CO + S + M \rightarrow OCS + M}
\end{equation}
\begin{equation}
    \mathrm{S + S_3 \rightarrow 2\,S_2}
\end{equation}
\begin{equation}
    \mathrm{H_2S + S \rightarrow 2\,HS}
\end{equation}
\begin{equation}
    \mathrm{S + S_4 \rightarrow S_2 + S_3}
\end{equation}

The reactions contributing to these variations in M4 vs M7 are shown in Figure~\ref{fig:ss2so_t_1e-6} for S, \ce{S2}, and SO. In M5 vs M7, the \ce{S2} evaporation process, $\mathrm{S_2(\ell) \rightarrow S_2(g)}$ (Figure~\ref{fig:ss2so_uv_1e-6}), emerges as an additional pathway not present in other model simulations. The condensation and evaporation of \ce{S2} are supported only in exo-Venus models that lack a stratospheric isotherm.

The abundance variations at mid-altitudes ($\sim$54~km) peak in the hybrid case M6 vs M9, followed by M4 vs M6, with minimal changes in M5 vs M7. These variations arise from differences in the formation and destruction rates of atomic sulfur between models with zero and non-zero (1~ppm) surface sulfur abundances. At these altitudes in M9, where excess sulfur from the surface is transported, the principal S production channels are:
\begin{equation}
    \mathrm{S_3 \xrightarrow{h\nu} S_2 + S}
\end{equation}
\begin{equation}
    \mathrm{S_4 \xrightarrow{h\nu} S_3 + S}
\end{equation}
\begin{equation}
    \mathrm{2\,SO \leftrightharpoons S + SO_2}
\end{equation}
\begin{equation}
    \mathrm{S_2O \xrightarrow{h\nu} SO + S}
\end{equation}
\begin{equation}
    \mathrm{S_2 \xrightarrow{h\nu} 2\,S}
\end{equation}

Figures~\ref{fig:ss2so_tuv_1e-6} and \ref{fig:h2socsco_tuv_1e-6} illustrate the reaction pathways for S, \ce{S2}, SO, \ce{H2S}, \ce{OCS}, and CO in M6 vs M9. As discussed previously, enhanced sulfur abundances modify the reaction kinetics of major sulfur production and loss channels, with effects propagating to higher-order polysulfur species through interconversion reactions. For \ce{S2}, the following destruction pathways contribute to polysulfur growth:
\begin{equation}
    \mathrm{2\,S_2 + M \leftrightharpoons S_4 + M}
\end{equation}
\begin{equation}
    \mathrm{S + S_2 + M \rightarrow S_3 + M}
\end{equation}
and additionally in M6:
\begin{equation}
    \mathrm{S_2 + S_6 + M \rightarrow S_8 + M}
\end{equation}

The OCS molecule also contributes to initiating polysulfur formation through reaction~\ref{eq:s_ocs--co_s2} in M9, which proceeds at rates several orders of magnitude higher than in M6. In contrast, M6 is dominated by:
\begin{equation}
    \mathrm{OCS \xrightarrow{h\nu} S + CO}
\end{equation}
Above $\sim$50~km, the rate of the excited-state channel~\ref{eq:ocs--s1d_co} in M9 remains slightly higher than that of the corresponding ground-state channel~\ref{eq:ocs--s_co}. For CO, the dominant reactions approximately balance each other through exchange with \ce{OCS}:
\begin{equation}
    \mathrm{CO + S + M \rightarrow OCS + M}
\end{equation}
\begin{equation}
    \mathrm{S + OCS \rightarrow CO + S_2}
\end{equation}
\begin{equation}
    \mathrm{OCS \xrightarrow{h\nu} S(^1D) + CO}
\end{equation}
\begin{equation}
    \mathrm{CO + HNO \rightarrow CO_2 + NH}
\end{equation}

In contrast to the Venusian cases, where 1~ppm sulfur primarily enhances \ce{S3} and \ce{S4} toward observational limits with minimal impact on other species (Figure~\ref{fig:venus_vmr}), the exo-Venus cases, particularly M6 vs M9, exhibit pronounced changes not only in polysulfur species but also in SO, \ce{H2S}, \ce{H2SO4}, \ce{S2O}, and CO at mid-altitudes. These results demonstrate that the presence of 1~ppm atomic sulfur near the surface can significantly reshape the chemical structure of exo-Venus atmospheres.

\section{Conclusions} \label{sec:conclusions}

In this work, we introduced the extended \texttt{XODIAC-2025} chemical network, \texttt{XODIAC-2025.v2}, for the fast, state-of-the-art 1D photochemical model \texttt{XODIAC} by incorporating excited-state sulfur chemistry and updated kinetic data suited to high-temperature, \ce{CO2}-dominated atmospheres. We applied this extended network to both Venus and exo-Venus analogs. Our key findings are summarized below:

\begin{enumerate}

\item The triplet S + CO$_2$ reaction proceeds through a two-step mechanism on the ground-state potential energy surface: an initiation step with a high entrance barrier of 284 kJ/mol, followed by a low-barrier ($<$10 kJ/mol) dissociation to products.

\item Two pathways were identified on the singlet excited-state potential energy surface for the S + CO$_2$ reaction: one involving a single transition state and another involving the crossing of two high-energy transition states.

\item We constructed a comprehensive reaction network that includes thermochemical and photochemical processes involving S, SO, S($^1$D), and $\mathrm{SO}(^1\Delta)$ to quantify their impact on the abundances of sulfur-bearing molecules in Venusian and exo-Venusian atmospheres.

\item To update the thermochemical database for the newly added species in \texttt{XODIAC-2025.v2}, such as S($^1$D) and $\mathrm{SO}(^1\Delta)$, we computed their NASA polynomial coefficients, which had not been previously reported. These coefficients were validated against those of ground-state S and SO from the Burcat thermochemical database.

\item We validated the base \texttt{XODIAC-2025.v1} network against the \texttt{STAND-2020} network for Venus by incorporating condensation processes and the empirical parameterization of the unknown Venusian UV absorber.

\item Comparison of the \texttt{XODIAC-2025.v1} and \texttt{XODIAC-2025.v2} networks for Venus shows that the added reactions introduce negligible differences below 60 km, with only minor photochemically driven deviations at higher altitudes. Both networks reproduce the observed abundances of major sulfur-bearing species well, except for \ce{S3} and \ce{S4}.

\item To assess the role of unaccounted sulfur sources on Venus, we compared two lower-boundary atomic sulfur conditions in \texttt{XODIAC-2025.v2}: zero sulfur and an imposed 1~ppm atomic sulfur abundance. Introducing 1~ppm S causes substantial increases in polysulfur abundances, particularly S$_3$ and S$_4$, bringing them into much closer agreement with observations. This result suggests the need for either a deeper, unmodeled source of reduced sulfur or more efficient polysulfur formation pathways than those presently included in existing networks.

\item We modeled three exo-Venus analogs, each with zero and 1~ppm initial atomic sulfur abundances: (i) a Venus-like $P$--$T$ structure with an isotherm above 30~km, (ii) an atmosphere exposed to 1000$\times$ solar flux, and (iii) a hybrid of both. Our exo-Venus models with initial $S = 0$ show only modest differences between the \texttt{XODIAC-2025.v1} and \texttt{XODIAC-2025.v2} networks in sulfur chemistry for the stratospheric isotherm model (M4), primarily due to photochemical branching in the production channels of S and S($^1$D).

\item Our exo-Venus models with initial $S = 1~\mathrm{ppm}$ (M7--M9) exhibit increasing deviations in sulfur abundances relative to the corresponding $S = 0$ cases (M4--M6), particularly in the lower and mid-atmospheres, driven primarily by enhanced sulfur production near the surface. The hybrid model M9 shows the largest deviations from M6 among all models, for both polysulfur species and most other sulfur-bearing molecules, reflecting the combined influence of strong irradiation and a high-altitude isotherm.

\item The $\mathrm{S + CO_2}$ reactions, involving both S($^3$P) and S($^1$D) and leading to the formation of \ce{SO(^3\Sigma)}, \ce{SO(^1\Delta)}, and \ce{CO} under \ce{CO2}-rich conditions, are kinetically unfavorable under Venusian and exo-Venusian environments due to their high activation barriers.

\end{enumerate}
These results highlight the need for more complete one-dimensional photochemical models that incorporate missing sources of deep-atmosphere sulfur and additional photochemical pathways leading to excited-state sulfur. Future work should focus on identifying and constraining these processes to improve agreement with Venusian observations and to enable more reliable predictions for exo-Venus analogs.
  
\section*{Acknowledgments}
L.M. also extends thanks to Sukrit Ranjan of the Lunar and Planetary Laboratory, Department of Planetary Sciences, University of Arizona, Tucson, USA, for insightful discussions on sulfur chemistry. P.G. thanks Alon Grinberg Dana for assistance with using WB97X-D/def2-TZVP level of theory in \texttt{RMG/Arkane}. L.M. acknowledges funding support from the DAE through the NISER project RNI 4011. L.M. also gratefully acknowledges support from Breakthrough Listen at the University of Oxford through a sub-award at NISER under Agreement R82799/CN013, provided as part of a global collaboration under the Breakthrough Listen project funded by the Breakthrough Prize Foundation.  Part of this research was conducted at the Jet Propulsion Laboratory, California Institute of Technology under contract with the National Aeronautics and Space Administration (80NM0018D004).  Support for K.W. was provided by NASA/Solar Systems Workings Program (19-SSW19-0291). N.R. thanks FONDECYT POSTDOCTORADO (ANID, Chile) grant 3230221 for financial support.



\newpage

\appendix

\section{T$_1$ diagnostic}

\begin{figure}[!h]
\includegraphics[width=\columnwidth]{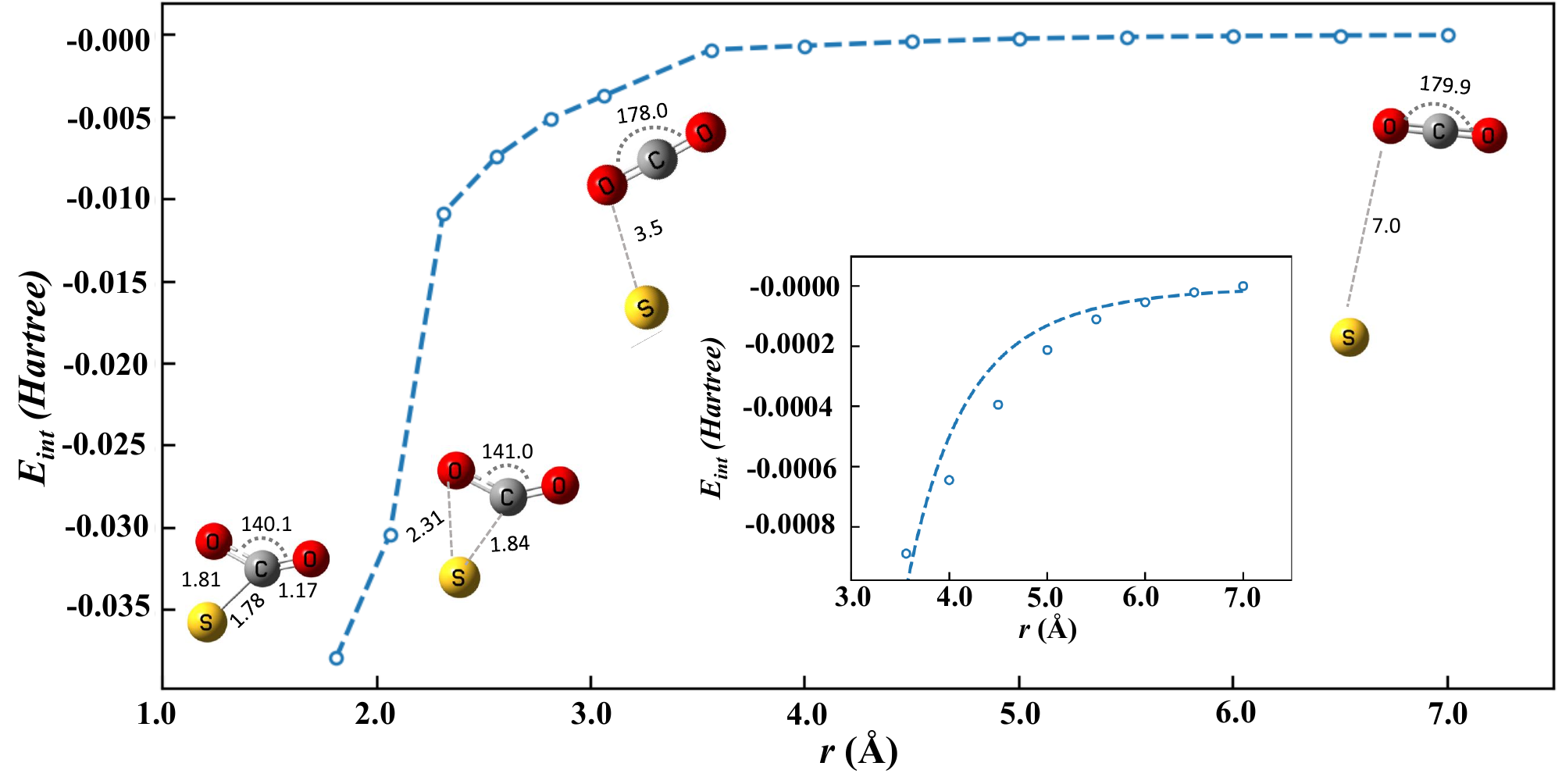}
\caption{Interaction-energy profile for the barrierless S + CO$_2$ entrance channel, Figure~\ref{fig:1}. Energies were obtained from a relaxed scan at the ROWB97X-D/def2-TZVP level with ROCCSD(T)/aug-cc-pV(T+d)Z single-point refinements. The relevant scan geometries with significant bond lengths (\AA) and bond angles (\textdegree) are depicted. All interaction energies are defined relative to the value at 7.0{~\AA}. The inset shows the long-range interaction energy fitted in region (3.5–7.0~\AA) to the $-C_6/R^6$ form. }

\label{fig:6}
\end{figure}

This section reports the T$_1$ diagnostic values obtained from CCSD(T) calculations for all singlet and triplet stationary points shown in Figure~\ref{fig:1} and listed in Table~\ref{tab:T1_diagnostics}. It also presents the interaction-energy scan used to model the long-range behavior of the barrierless \ce{S + CO2} entrance channel (Figure~\ref{fig:6}).

\begin{table}[h!]
\centering
\caption{T$_1$ diagnostic from CCSD(T) calculations for all singlet and triplet stationary points from Figure~\ref{fig:1}.}
\label{tab:T1_diagnostics}
\begin{tabular}{cc}
\hline\hline
Stationary point & T$_1$ diagnostic \\
\hline
$^{3}$RC     & 0.028 \\
$^{3}$TS$_1$ & 0.022 \\
$^{3}$TS$_2$ & 0.027 \\
\hline
$^{1}$RC     & 0.019 \\
$^{1}$int$_1$ & 0.023 \\
$^{1}$TS     & 0.024 \\
$^{1}$TS$_1$ & 0.029 \\
$^{1}$TS$_2$ & 0.027 \\
\hline\hline
\end{tabular}
\end{table}

\clearpage

\section{Major Reaction Pathways for Venusian and exo-Venusian model simulations}

\begin{figure*}
    \centering
    \includegraphics[width=0.9\linewidth]{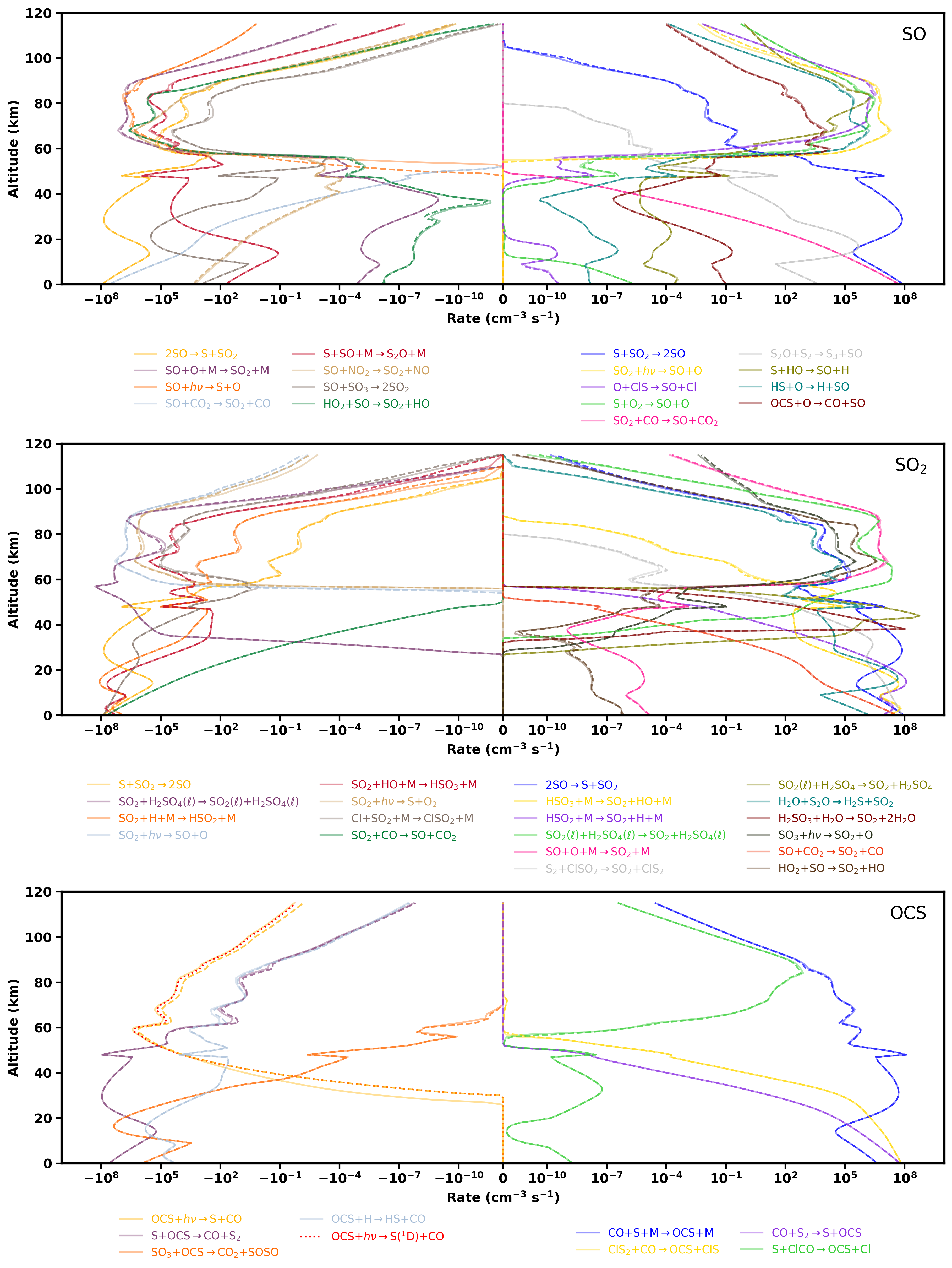}
  \caption{Rates of the major reactions with altitude in the Venus atmosphere for \ce{SO}, \ce{SO2}, and \ce{OCS}. Solid and dashed lines correspond to results from the base \texttt{XODIAC-2025.v1} network (M0) and the newly developed \texttt{XODIAC-2025.v2} network (M2), respectively. Reactions whose labels appear in dotted style (if any) are contributing from the \texttt{XODIAC-2025.v2} network only. Negative rate values indicate that the reaction acts as a sink for the species, whereas positive values indicate a source. Symmetrical values on opposite sides of the zero rate for any reaction imply equal rates of formation and destruction. Reactions contributing less than 10\% are omitted for clarity.}

    \label{fig:soso2h2s_m0m2}
\end{figure*}

\begin{figure*}
    \centering
    \includegraphics[width=0.9\linewidth]{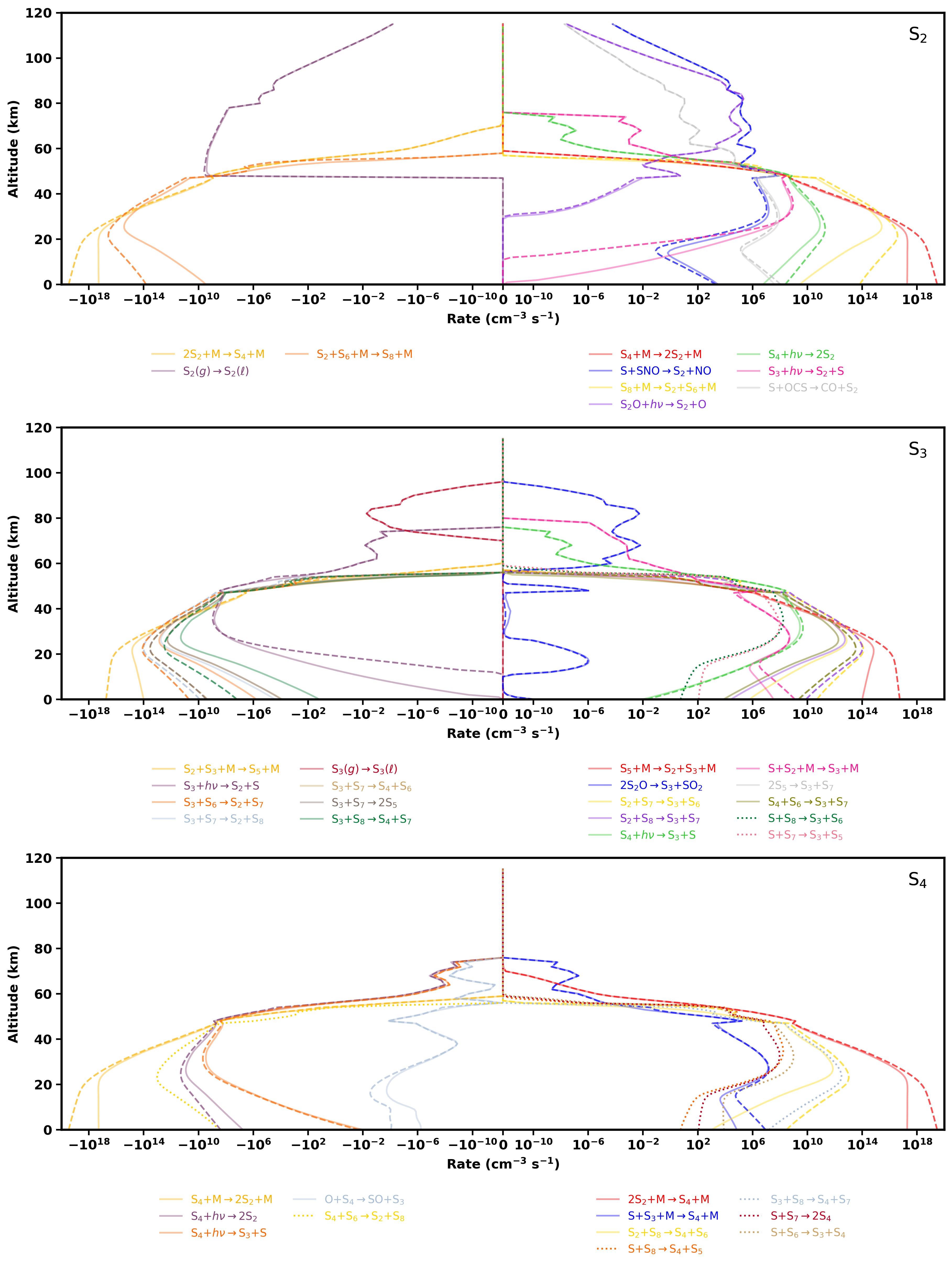}
    \caption{Rates of the major reactions with altitude in the Venus atmosphere for \ce{S2}, \ce{S3} and \ce{S4}. Solid and dashed lines correspond to results from base \texttt{XODIAC-2025.v1} network (M0) and \texttt{XODIAC-2025.v1} with 1~ppm atomic S (M1), respectively. Reactions whose labels appear in dotted style (if any) are contributing from M1 only. Negative rate values indicate that the reaction acts as a sink for the species, whereas positive values indicate a source. Symmetrical values on opposite sides of the zero rate for any reaction imply equal rates of formation and destruction. Reactions contributing less than 10\% are omitted for clarity.} 
    \label{fig:s2s3s4_m0m1}
\end{figure*}

\begin{figure*}
    \centering
    \includegraphics[width=0.9\linewidth]{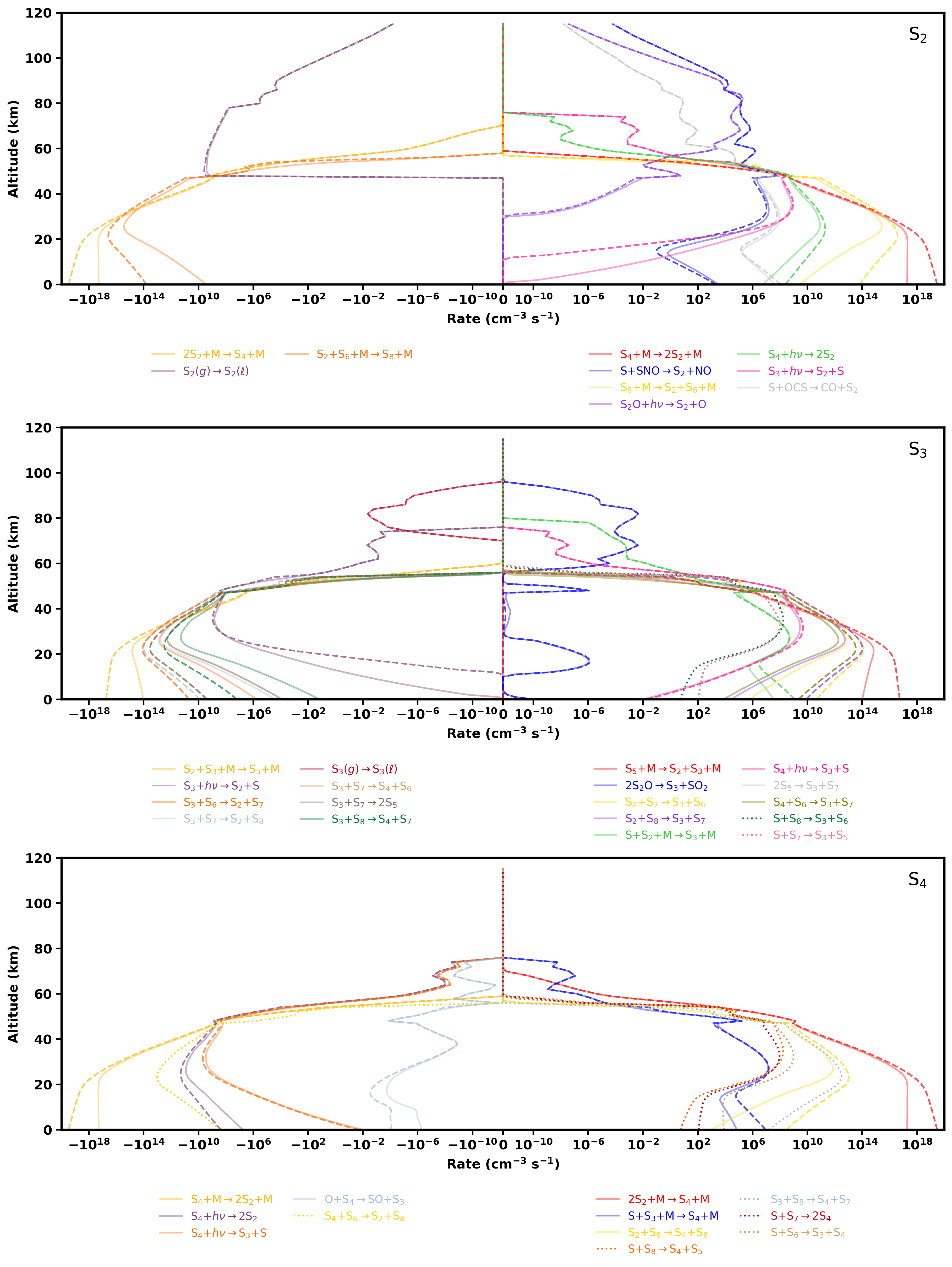}
    \caption{Rates of the major reactions with altitude in the Venus atmosphere for \ce{S2}, \ce{S3} and \ce{S4}. Solid and dashed lines correspond to results from the newly developed \texttt{XODIAC-2025.v2} network (M2) and \texttt{XODIAC-2025.v2} with 1~ppm atomic S (M3), respectively. Reactions whose labels appear in dotted style (if any) are contributing from M3 only. Negative rate values indicate that the reaction acts as a sink for the species, whereas positive values indicate a source. Symmetrical values on opposite sides of the zero rate for any reaction imply equal rates of formation and destruction. Reactions contributing less than 10\% are omitted for clarity.}   
    \label{fig:s2s3s4_m1m2}
\end{figure*}

\begin{figure*}
    \centering
    \includegraphics[width=0.9\linewidth]{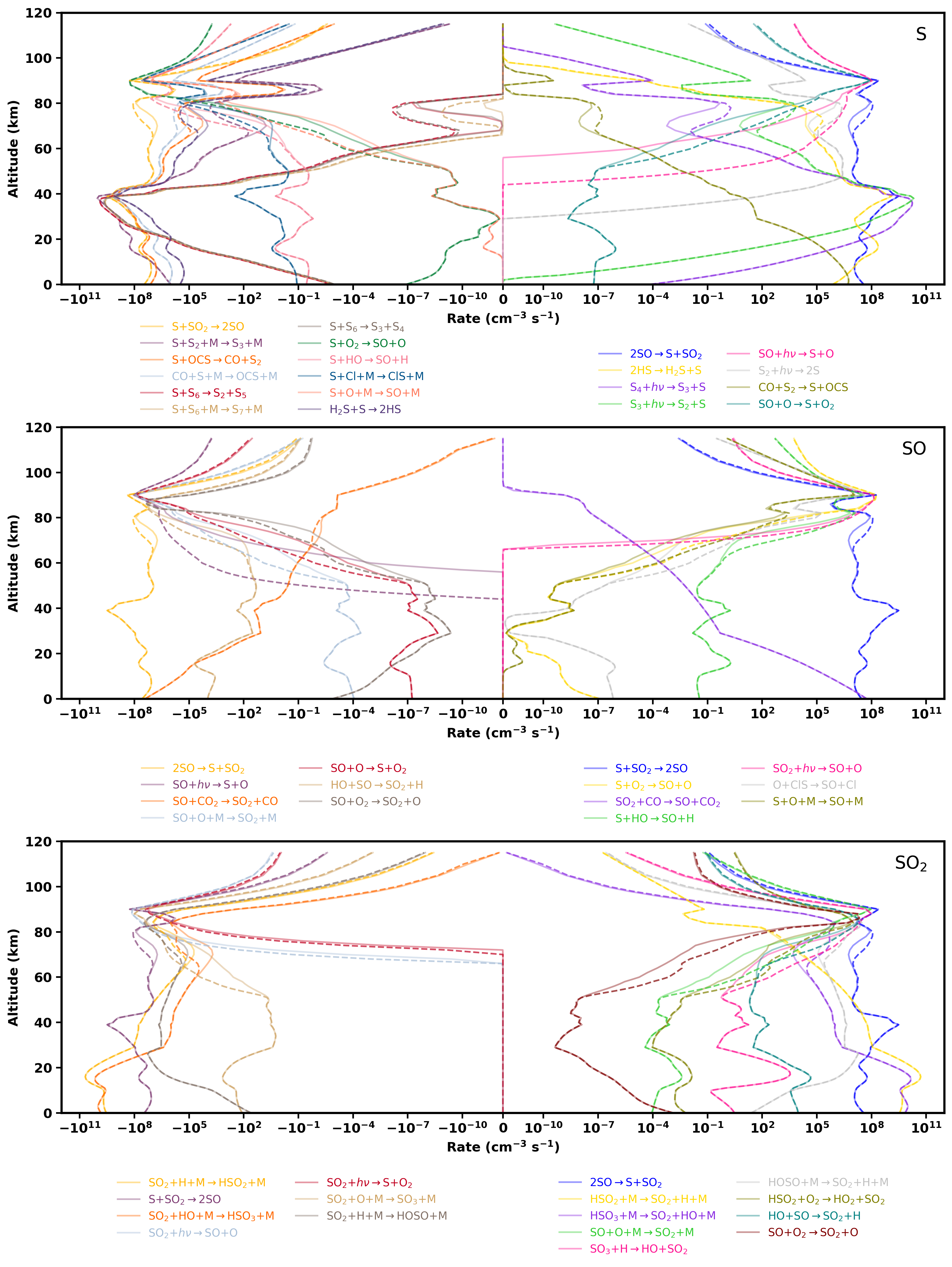}
    \caption{Rates of the major reactions as a function of altitude in the exo-Venus atmosphere (Model M4) for S, \ce{SO}, and \ce{SO2}. Solid and dashed lines correspond to results from the base \texttt{XODIAC-2025.v1} network and the newly developed \texttt{XODIAC-2025.v2} network, respectively. Reactions whose labels appear in dotted style (if any) are contributing from the \texttt{XODIAC-2025.v2} network only. Negative rate values indicate that the reaction acts as a sink for the species, whereas positive values indicate a source. Symmetrical values on opposite sides of the zero rate for any reaction imply equal rates of formation and destruction. Reactions contributing less than 10\% are omitted for clarity.}
    \label{fig:ssoso2_m4}
\end{figure*}

\begin{figure*}
    \centering
    \includegraphics[width=0.9\linewidth]{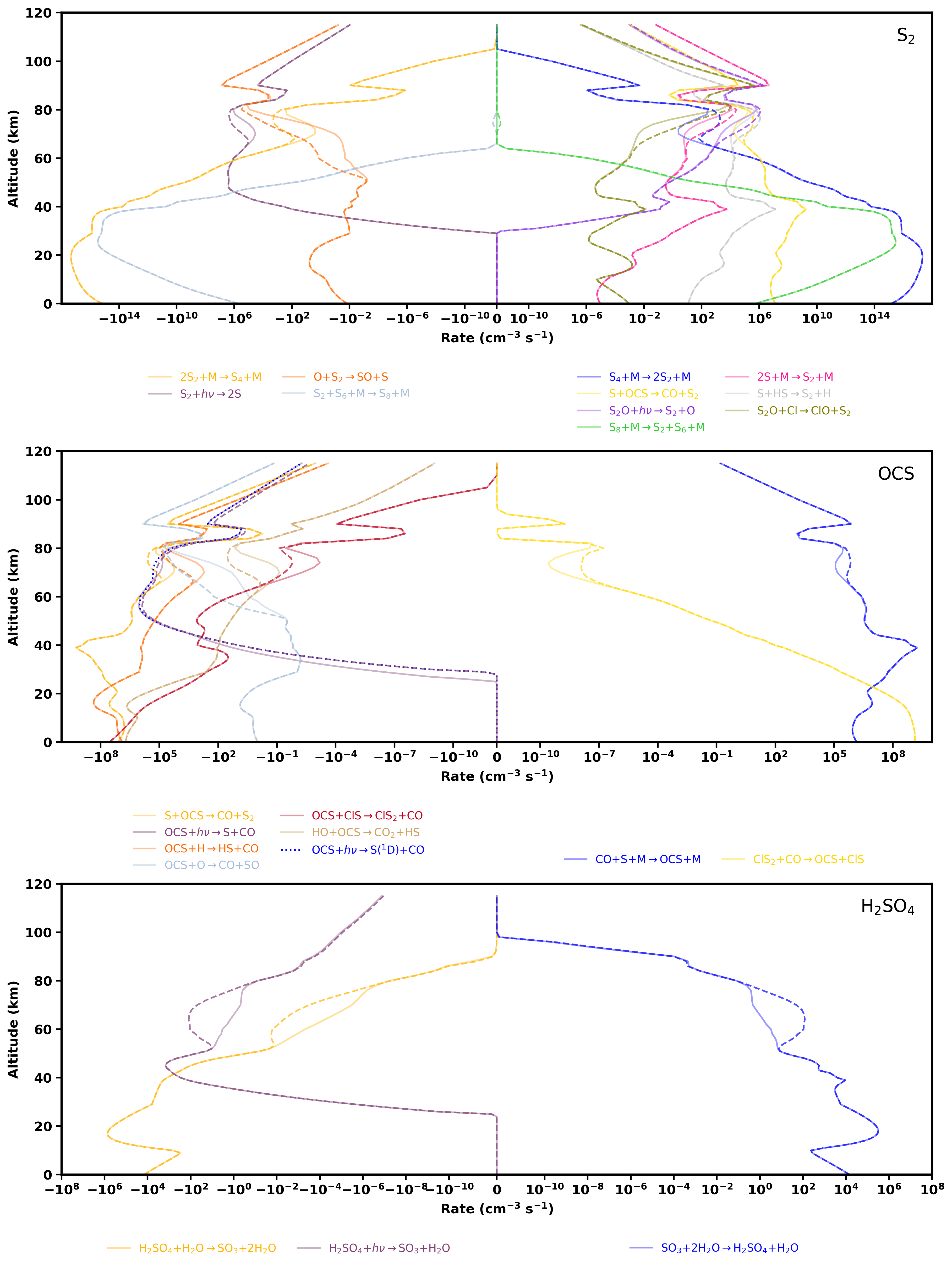}
    \caption{Same as Figure \ref{fig:ssoso2_m4}, but for the exo-Venus atmosphere (Model M4), showing the major reaction rates for \ce{S2}, \ce{OCS}, and \ce{H2SO4 (g)}.}
    \label{fig:s2ocsh2so4_m4}
\end{figure*}

\begin{figure*}
    \centering
    \includegraphics[width=0.9\linewidth]{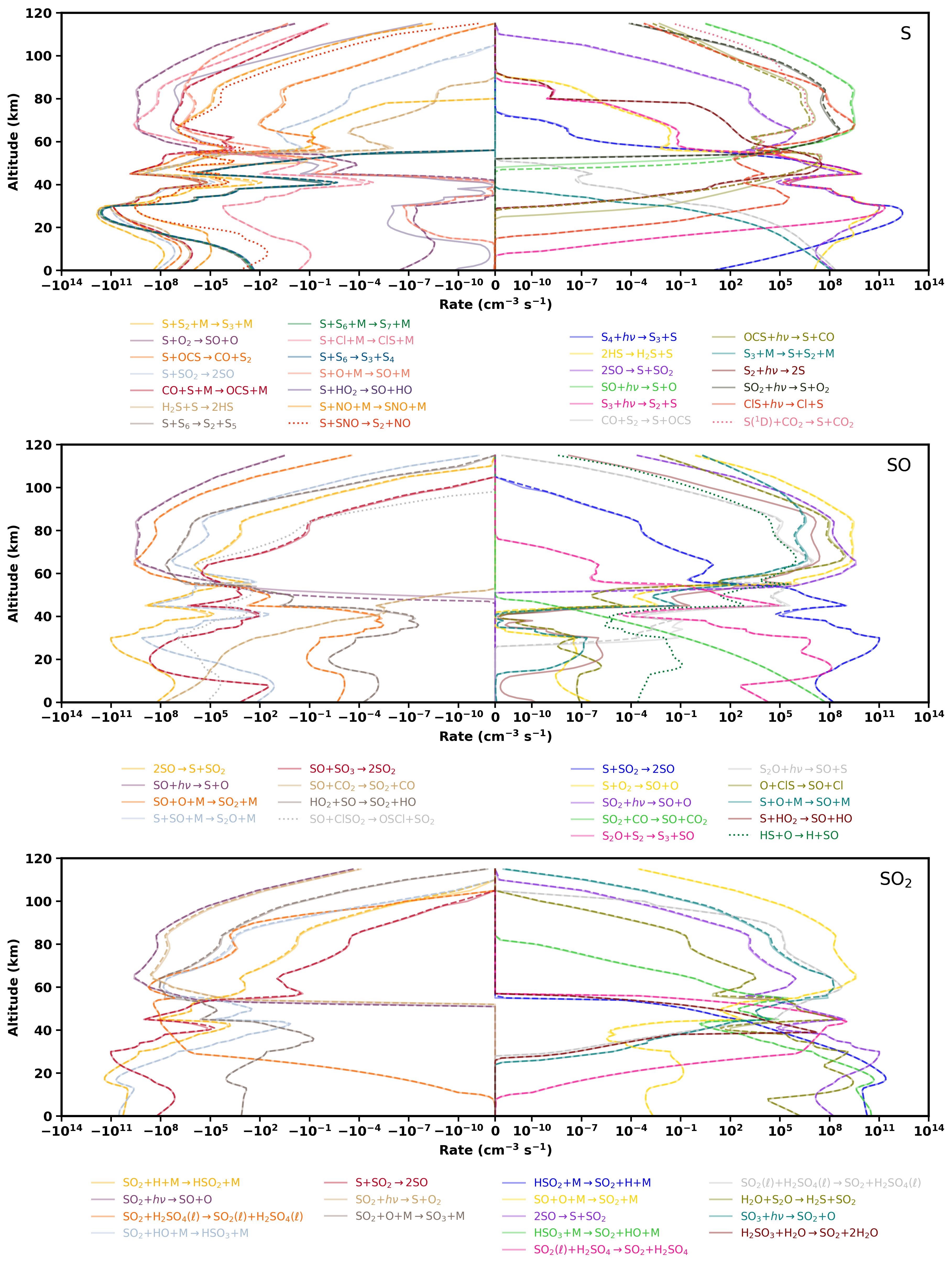}
    \caption{Same as Figure~\ref{fig:ssoso2_m4}, but for the exo-Venus atmosphere (Model M5), showing the major reaction rates for S, SO, \ce{SO2}.}    
    \label{fig:ssoso2_m5}
\end{figure*}

\begin{figure*}
    \centering
    \includegraphics[width=0.9\linewidth]{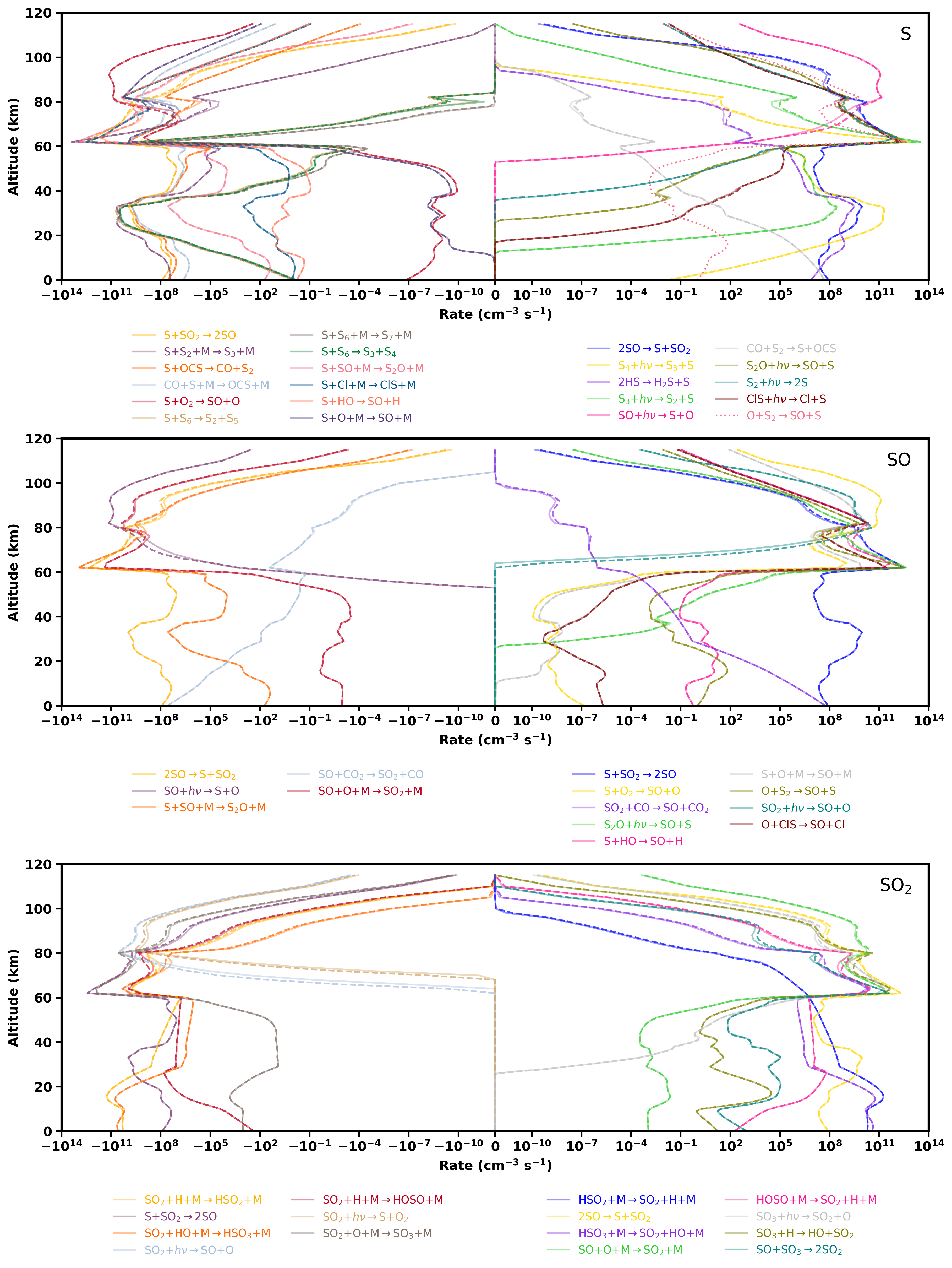}
    \caption{Same as Figure \ref{fig:ssoso2_m4}, but for the exo-Venus atmosphere (Model M6), showing the major reaction rates for S, SO, \ce{SO2}.}

    \label{fig:ssoso2_m6}
\end{figure*}

\begin{figure*}
    \centering
    \includegraphics[width=0.9\linewidth]{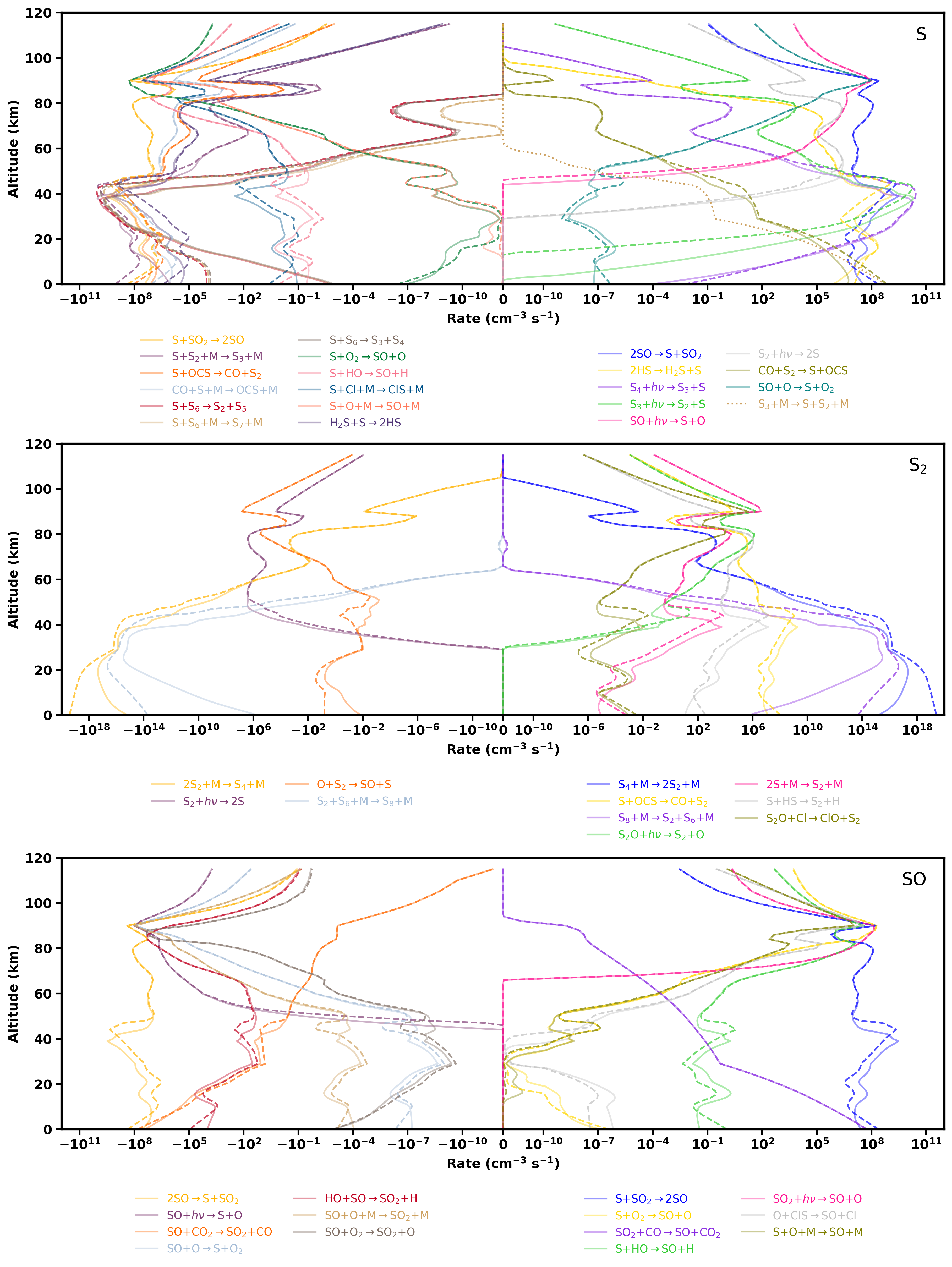}
    \caption{Rates of the major reactions as a function of altitude in the exo-Venus atmosphere Model M4 vs Model M7 for S, \ce{S2}, and \ce{SO}. Solid and dashed lines correspond to results from the \texttt{XODIAC-2025.v2} using S = 0 and S = 1 ppm, respectively. Reactions whose labels appear in dotted style (if any) are contributed by Model M7 only. Negative rate values indicate that the reaction acts as a sink for the species, whereas positive values indicate a source. Symmetrical values on opposite sides of the zero rate for any reaction imply equal rates of formation and destruction. Reactions contributing less than 10\% are omitted for clarity.}
    
    \label{fig:ss2so_t_1e-6}
\end{figure*}

\begin{figure*}
    \centering
    \includegraphics[width=0.9\linewidth]{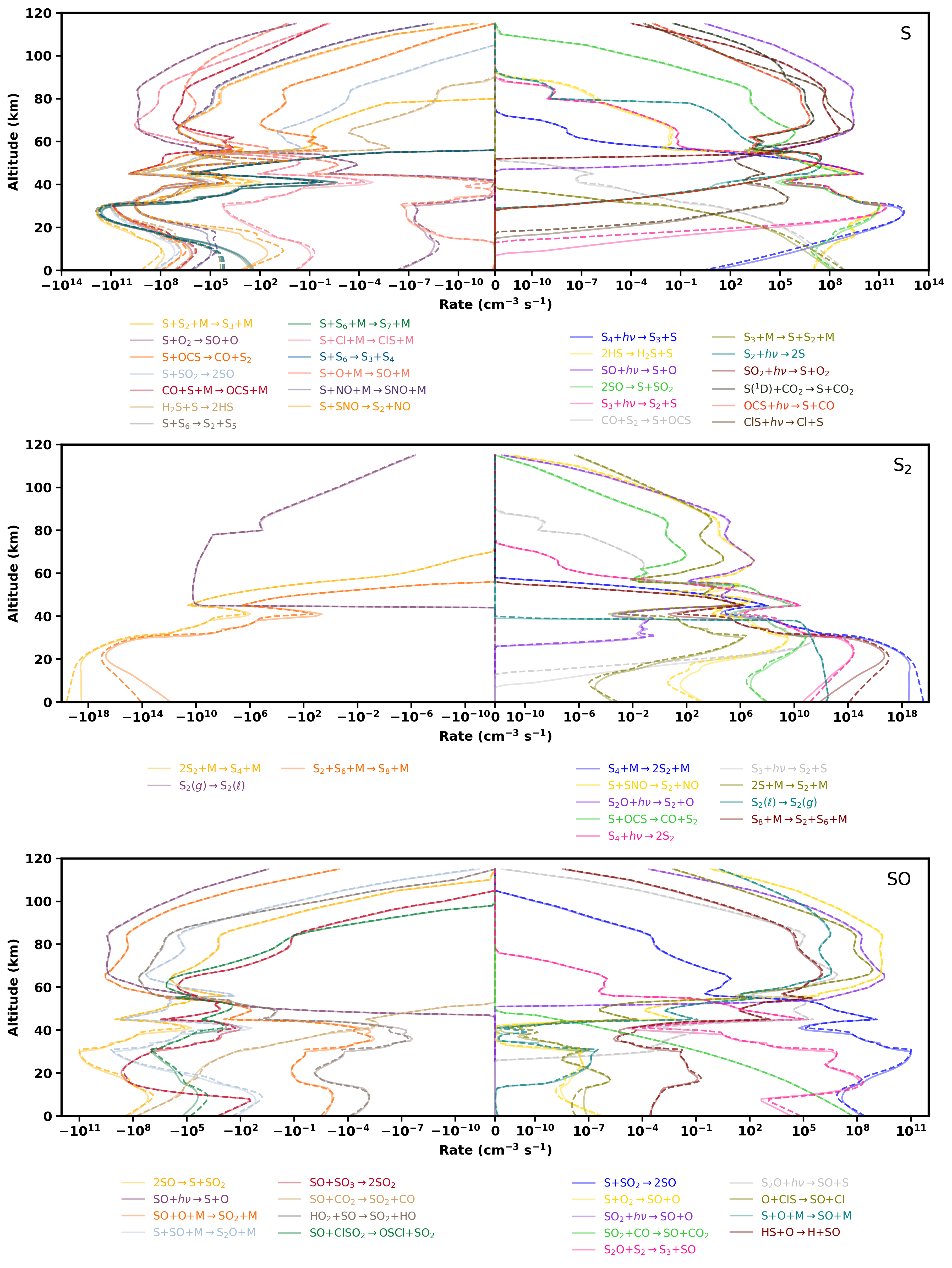}
    \caption{Same as Figure \ref{fig:ss2so_t_1e-6}, but for the exo-Venus atmosphere Model M5 vs Model M8, showing the major reaction rates for S, \ce{S2}, and \ce{SO}. Reactions whose labels appear in dotted style (if any) are contributed by Model M8 only.}
    \label{fig:ss2so_uv_1e-6}
\end{figure*}

\begin{figure*}
    \centering
    \includegraphics[width=0.9\linewidth]{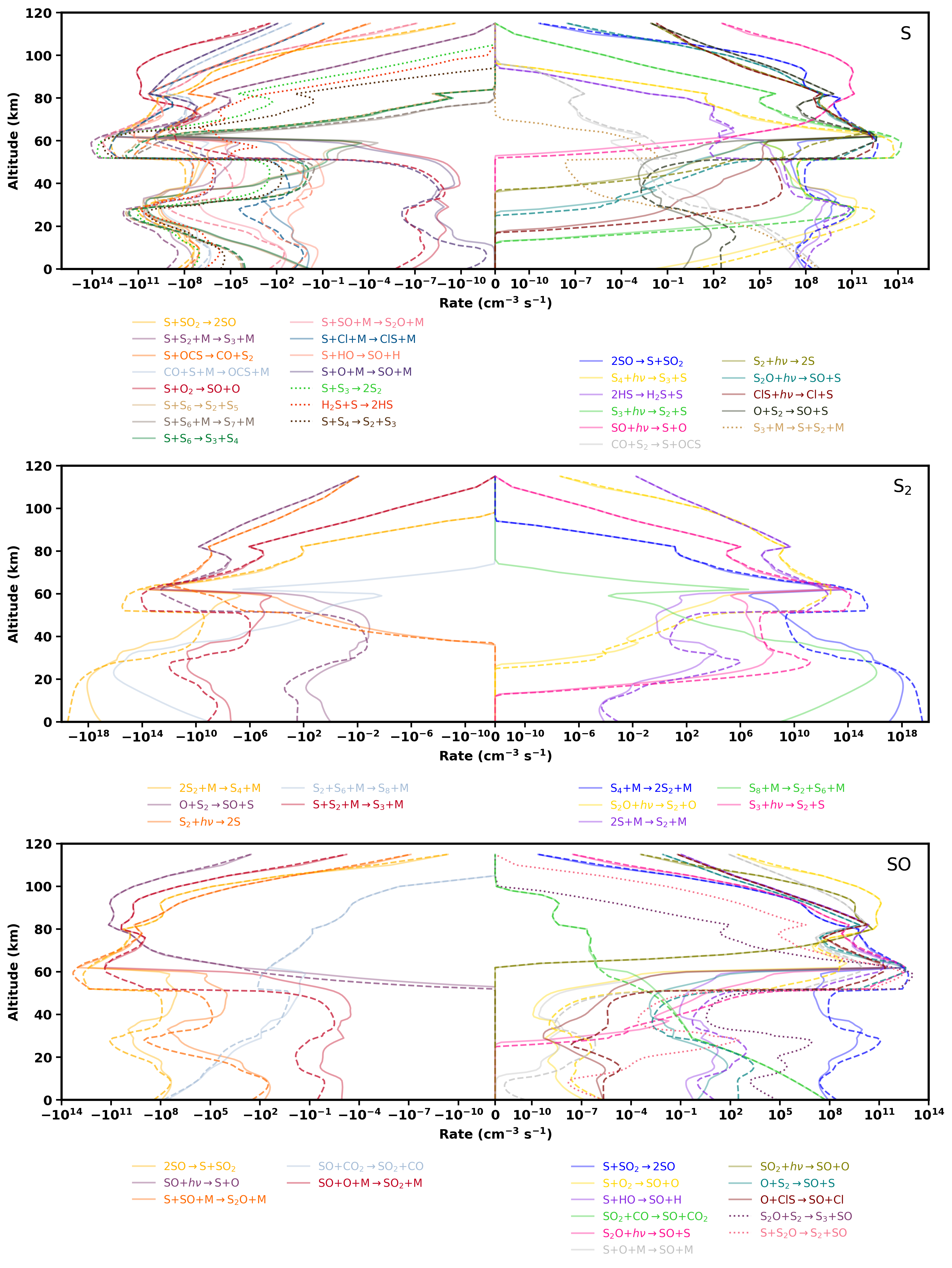}
    \caption{Same as Figure \ref{fig:ss2so_t_1e-6}, but for the exo-Venus atmosphere Model M6 vs Model M9, showing the major reaction rates for S, \ce{S2}, and \ce{SO}. Reactions whose labels appear in dotted style (if any) are contributed by Model M9 only.}
    \label{fig:ss2so_tuv_1e-6}
\end{figure*}

\begin{figure*}
    \centering
    \includegraphics[width=0.9\linewidth]{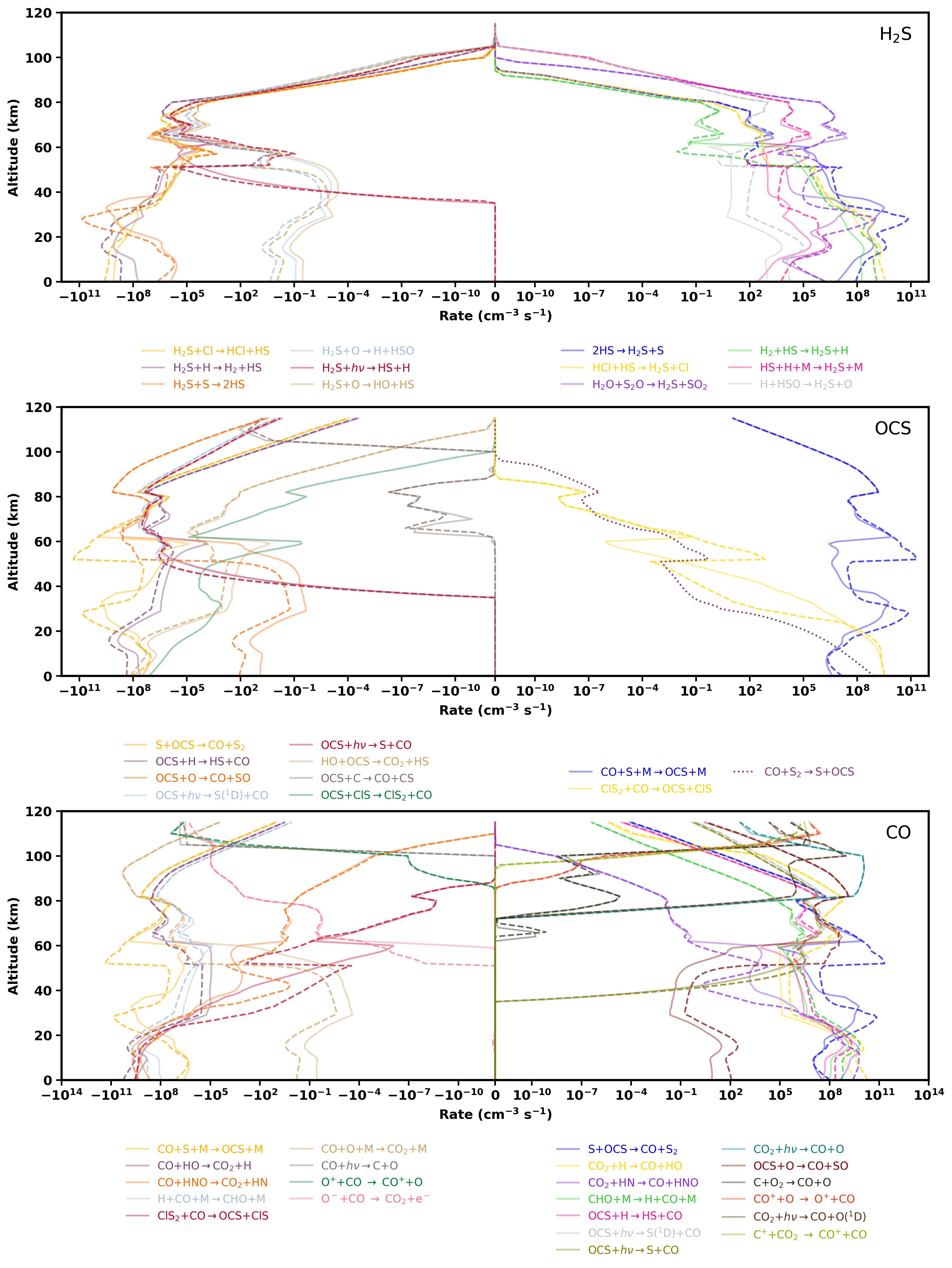}
    \caption{Same as Figure \ref{fig:ss2so_t_1e-6}, but for the exo-Venus atmosphere Model M6 vs Model M9, showing the major reaction rates for \ce{H2S}, \ce{OCS}, and CO. Reactions whose labels appear in dotted style (if any) are contributed by Model M9 only.}
    \label{fig:h2socsco_tuv_1e-6}
\end{figure*}

In this section, the profiles of the major reaction rates are presented for all models (M0--M9) relevant to Venus and exo-Venus atmospheric chemistry in Figures~\ref{fig:soso2h2s_m0m2} to \ref{fig:h2socsco_tuv_1e-6}. For the Venusian models, Figure~\ref{fig:soso2h2s_m0m2} shows \ce{SO}, \ce{SO2}, and \ce{OCS}, as their abundances exhibit the largest variations between M0 and M2 (with no polysulfur variations). Figures~\ref{fig:s2s3s4_m0m1} and \ref{fig:s2s3s4_m1m2} illustrate the effects of adding 1~ppm atomic sulfur and the associated polysulfur chemistry involving \ce{S2}, \ce{S3}, and \ce{S4}, while maintaining internal consistency within each network.

For the exo-Venusian models, Figures~\ref{fig:ssoso2_m4} and \ref{fig:s2ocsh2so4_m4} illustrate variations in S, SO, \ce{SO2}, \ce{S2}, OCS, and \ce{H2SO4} induced by the imposed stratospheric isotherm, which results in modest deviations. Figure~\ref{fig:ssoso2_m5} shows the response of S, SO, and \ce{SO2} to enhanced stellar irradiation. Figure~\ref{fig:ssoso2_m6} presents the reaction pathways involving S, SO, and \ce{SO2} for the hybrid case combining a stratospheric isotherm and enhanced stellar irradiation. The major reaction rates for $S = 0$ and $S = 1~\mathrm{ppm}$ for the three exo-Venus cases (stratospheric isotherm, strong stellar irradiation, and their hybrid) are shown in Figures~\ref{fig:ss2so_t_1e-6}, \ref{fig:ss2so_uv_1e-6}, and \ref{fig:ss2so_tuv_1e-6}, respectively, for S, \ce{S2}, and SO. Additional variations in \ce{H2S}, OCS, and CO (for M6 vs M9) are shown in Figure~\ref{fig:h2socsco_tuv_1e-6}, as these species exhibit the largest changes.


\clearpage

\bibliography{References}

\bibliographystyle{aasjournal}

\end{document}